\documentclass[fleqn,usenatbib]{mnras}

\usepackage{newtxtext,newtxmath}

\usepackage[T1]{fontenc}

\DeclareRobustCommand{\VAN}[3]{#2}
\let\VANthebibliography\thebibliography
\def\thebibliography{\DeclareRobustCommand{\VAN}[3]{##3}\VANthebibliography}

\usepackage{graphicx}	
\usepackage{amsmath}	
\usepackage{multirow}

\usepackage{orcidlink}

\title[Cloud cover and meteorology at Lenghu site]{The cloud cover and meteorological parameters at the Lenghu site on the Tibetan Plateau}

\author[R, Y, Li et al.]{
Ruiyue Li,$^{1,2}$
Fei He,$^{3,4}$
Licai Deng\orcidlink{0000-0001-9073-9914},$^{1,4}$\thanks{E-mail: licai@nao.cas.cn(LD)} 
Xiaodian Chen,$^{1,4}$
Fan Yang,$^{1,4}$
Yong Zhao,$^{3}$
Bo Zhang,$^{1}$
\newauthor{
Chunguang Zhang,$^{1}$
Chen Yang,$^{4,1}$
Tian Lan,$^{5,1}$}
\\
$^{1}$National Astronomical Observatories, Chinese Academy of Sciences, 100101 Beijing, China\\
$^{2}$School of Astronomy and Space Science, University of Chinese Academy of Sciences, 100049 Beijing, China\\
$^{3}$Institute of Geology and Geophysics, Chinese Academy of Sciences, 100029 Beijing, China\\
$^{4}$School of Physics and Astronomy, West-China Normal University, 637001 Nanchong, China\\
$^{5}$Department of Physics, College of Science, Tibet University, 850000 Lhasa, China}

\date{Accepted XXX. Received YYY; in original form ZZZ}

\pubyear{2015}

\begin{document}
\label{firstpage}
\pagerange{\pageref{firstpage}--\pageref{lastpage}}
\maketitle

\begin{abstract}

The cloud cover and meteorological parameters serve as fundamental criteria for the qualification of an astronomical observatory working in optical and infrared wavelengths. In this paper, we present a systematic assessment of key meteorological parameters at Lenghu, a high-quality observing site on the Tibetan Plateau. The datasets adopted in this study come from a number of sources based on different techniques, including the meteorological parameters collected at the local weather stations at the site and in the Lenghu Town, the sky brightness at the local zenith acquired by the Sky Quality Meters and night sky all-sky images from a digital camera, the ERA5 reanalysis database and global climate monitoring data. From 2019 to 2023, the fractional observable time of photometric condition is 69.70\%, 74.97\%, 70.26\%, 74.27\% and 65.12\%, respectively. The fractional observing time is inversely correlated with surface air temperature, relative humidity, precipitable water vapor, and dew temperature, demonstrating that the observing conditions are influenced by these meteorological parameters. Large-scale air-sea interactions affect the climate at Lenghu site, which in fact delivers a clue to understand the irregularity of 2023. Specifically, precipitable water vapor at Lenghu site is correlated to both the westerly wind index and the summer North Atlantic Oscillation index, the yearly average temperature of Lenghu site is observed to increase significantly during the occurrence of a strong El Niño event and the relative humidity anomaly at Lenghu site is correlated to the Pacific Decadal Oscillation index. The decrease of fractional observing time in 2023 was due to the ongoing strong El Niño event and relevant global climate change. We underscore the substantial role of global climate change in regulating astronomical observing conditions and the necessity for long-term continuous monitoring of the astronomical meteorological parameters at Lenghu site.

\end{abstract}

\begin{keywords}
Site testing -- telescopes -- methods: data analysis -- atmospheric effects
\end{keywords}



\section{Introduction}

The Lenghu astronomical observing site is located at a local summit of Saishiteng Mountain near the Lenghu town, which sits at elevations of 4200 to 4400 meters. It belongs to the western branch of the Qilian Mountains, the northern edge of the Qaidam Basin and the east of the Altyn Mountains \citep{deng2021lenghu, pullen2011qaidam}. The average elevation of Qaidam basin is around 2700 meters and the Saishiteng Mountain rises abruptly from it, forming a relative elevation of nearly 1,500 meters in only 20 km. On a global scale, the site is in the hinterland of Eurasia and has a great distance from the ocean. Under the impact of the subtropical high and the special landform of the Tibetan Plateau, it is one of the most arid regions in the world. As a typical area of highland climate, Lenghu site also has the climate characteristic of high insolation, low temperature and low air pressure. It is the exact geographic location of extremely dry and cloudless air conditions at high altitudes that gives Lenghu a great potential to host large astronomical facilities working in optical and infrared wavelengths.

The local climate conditions serve as a key factor in the observing quality of ground-based optical/IR telescopes. Longest possible observing time (i.e. lowest possible cloud cover) and finest astronomical seeing are what we need during a site survey, the associated important atmospheric parameters are worth monitoring \citep{schock2009thirty}. In order to host modern large aperture telescopes and to apply extreme techniques in cutting-edge astrophysical problems, the most critical factors are the meteorological conditions including surface air temperature, surface air pressure, relative humidity and wind speed, and of course the cloud cover statistics \citep{schock2009thirty, vernin2011european, seidel2023impact}. In addition, astronomers are very much concerned about the integrated absolute humidity: precipitable water vapour (PWV), which blocks near-to-mid infrared radiations from celestial bodies due to absorption and thermal emissions \citep{bustos2014parque}. In Sect.~\ref{sec:result}, we compiled the long-term trends of all the aforementioned parameters and analyzed their seasonal characteristics.

The climate data collected in the last few years since 2018 at the Lenghu site are favorable \citep{deng2021lenghu}, but when implementing the planning of next-generation telescopes, from site selection to end-of-life span which often covers at least 30 to 50 years, we hope to find the annual and decadal variations of the site parameters so that we can draw the blueprint for long-term development of the site and maximize its scientific potential. Knowing the general characteristics of the site, the community still needs to learn more about the average trends beyond seasonal variations, especially the impact of large-scale or even global climate changes on future scientific operations. Especially in infrared wavelengths that are critical for cutting-edge sciences \citep{seidel2023impact}. The annual and decadal trends of regional climate conditions cannot be separated from the large-scale/global atmospheric circulation and ocean-atmosphere climate variabilities. In this paper, we try to find the relationship between the meteorological parameters of the Lenghu site with the global pictures. The climate pattern of the Tibetan Plateau is subject to a variety of air-sea interactions, among which we found that the site's PWV is correlated with the North Atlantic Oscillation (NAO) index and the regional westerly index, the relative humidity (RH) measured at the site is associated with Pacific Decadal Oscillation (PDO). We observed a significant increase in the annual average temperature at the Lenghu site during the occurrence of a strong El Niño event. In 2023, there was a notable increase in the overall annual average temperature. The discussion is given in Sect.~\ref{sec:discussion}.

\begin{figure}
    \centering
	\includegraphics[width=0.95\columnwidth]{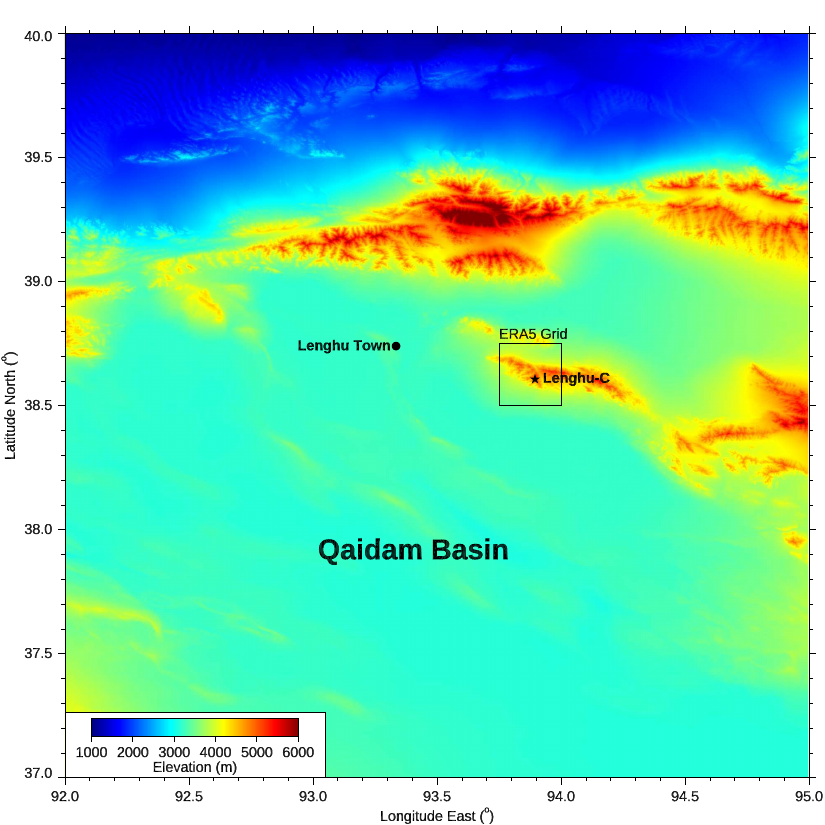}
    \caption{Elevation map of Lenghu region. The elevation data are downloaded from the AW3D of the Japan Aerospace Exploration Agency (JAXA) with a spatial resolution of 1 arcsec (30 meters) and an elevation accuracy of 1 meter. The black star denotes the location of Lenghu-C at the Saishiteng Mountain. The location of Lenghu Town is marked by the black dot. The black rectangle marks the grid size of ERA5 (latitude 38.5º to 38.75º, longitude 93.75º to 94.0º). The Lenghu-C is approximately located at the center of this grid. The color bar for the elevation is shown at the lower left corner. The distance between Lenghu-C and Lenghu Town is 50.8 km.}
    \label{topographic}
\end{figure}

\section{Data} 

In this work, three astronomical meteorological datasets and key global climate monitoring data are adopted to systematically investigate the characteristics of Lenghu site. Namely, ERA5 \citep{Hersbach2020}, the latest generation of atmospheric reanalysis of European Centre for Medium-Range Weather Forecasts (ECMWF), the data collected at local weather stations on site for site studies, and the historic weather data from Haixi Meteorological Administration. Table~\ref{tab:data} summarizes all the datasets used in this paper, including information on archive length, data source, temporal resolution and spatial resolution. Each dataset is followed by a description that explains the purpose of the data analysis in this study.

ERA5 offers hourly data on numerous climate variables related to atmosphere, land, and ocean. The dataset has a horizontal grid of 0.25$^\circ$ $\times$ 0.25$^\circ$, which equals to a spatial resolution of approximately 30 kilometers. The earth's atmosphere is stratified into 37 vertical levels, extending from the Earth's surface to an altitude of 80 kilometers. We adopted the ERA5 data with the nearest horizontal grid of Lenghu-C (38.6068$^\circ$N, 93.8961$^\circ$E), which is 38.5$^\circ$N and 94.0$^\circ$E and is illustrated by the black rectangle in Fig.~\ref{topographic}. The ERA5 reanalysis data enables us to do the long-term analysis. In this paper, we obtain the dataset from 2000 to 2023 (full 24 years), which covers two solar cycles, to study the long-term variations at Lenghu site.

The data we used from ground weather station of Lenghu-C were collected at the Saishiteng Mountain which is marked by the black star in Fig.~\ref{topographic}. With an elevation of ~4200 m, the site parameters were measured at a pressure level of approximately 607 hPa. The time resolution is one minute, and it provides 5 meteorological parameters, including wind speed and direction, temperature, relative humidity and air pressure from 2018.10 to 2023.12 (more than five years).

The third part of the atmospheric data is provided by the Haixi Meteorological Administration. This part of data was taken at the standard weather station at Lenghu Town from 2005.1 to 2017.9 (more than 12 years) which is also marked by the black dot in Fig.~\ref{topographic}. Lenghu Town has an elevation of 2733 m and is 50.8 km away from the Lenghu site. As the nearest populated area to the site, Lenghu Town is crucial for providing logistic material and transportation, so it is necessary to compare the atmospheric parameters of Lenghu Town with the ERA5 data of the Lenghu site.

We also use key global climate monitoring data (Niño 3.4 index, PDO index and NAO index) for site analysis, which can be accessed from the website of National Oceanic and Atmospheric Administration: National Centers for Environmental Information.The local climate pattern of an astronomical site plays a crucial role in determining the observation conditions, specifically the observable time. This pattern is primarily influenced by the geographical location, particularly the seasonal variations. However, it is important to note that annual and interannual variations are also impacted by global climate change. Lenghu site is just beneath the southern edge of the westerly in the northern hemisphere, which is apparently affected by global climate change. Therefore, investigating the correlation between climate parameters at Lenghu site and key global climate indices may yield valuable insights into the long-term variation of the observable time at Lenghu site.

\begin{table*}
	\centering
    \setlength{\tabcolsep}{4pt}
	\caption{A detailed table summarizing all the datasets used in this paper, including information on archive length, temporal resolution and spatial resolution, data source is shown. Each dataset is accompanied by a description that explains the purpose of the data analysis in this study.}
	\label{tab:data}
	\begin{tabular}{cccccc} 
		\hline
		\multirow{2}{*}{Dataset} & \multirow{2}{*}{Archive length} & \multicolumn{2}{c|}{Resolution} & \multirow{2}{*}{Data source} \\
        & & Temporal & Spatial \\
		\hline
		ERA5 & 2000.1-2023.12 & 1 hour & 0.25$^\circ$ (\textasciitilde 30 km) & \url{https://cds.climate.copernicus.eu/cdsapp} \\
        & \multicolumn{4}{l}{\textit{Description: This dataset is used to study the long-term trends of meteorological parameters at the Lenghu site, which}} \\
        & \multicolumn{4}{l}{\textit{helps us understand the local climate patterns.}} \\
        \hline
        Lenghu-C & 2018.10-2023.12 & 1 minute & single point & ground weather station \url{http://lenghu.china-vo.org/} \\
        & \multicolumn{4}{l}{\textit{Description: In-situ cloud cover and meteorological parameters measured by Lenghu-C ground weather station.}} \\
        \hline
        Lenghu Town & 2005.1-2017.9 & 1 hour & single point & Haixi Meteorological Administration \\
        & \multicolumn{4}{l}{\textit{Description: Lenghu Town is the nearest populated area to the site and crucial for providing logistic material and}}\\
        & \multicolumn{4}{l}{\textit{transportation, this dataset is used to compare the atmospheric parameters with that of the Lenghu site.}} \\
        \hline
        Global Climate Monitoring Data & 2000.1-2023.12 & 1 month & & \url{https://www.ncei.noaa.gov/access/monitoring/products/}\\
        & \multicolumn{4}{l}{\textit{Description: investigating the correlation between in-situ meteorological parameters and key global climate indices may}} \\
        & \multicolumn{4}{l}{\textit{yield valuable insights into the long-term variation of the observable time at Lenghu site.}} \\
        \hline
	\end{tabular}
\end{table*}

\section{Results}
\label{sec:result}

\subsection{Observable Time}

\begin{figure}
    \centering
	\includegraphics[width=0.95\columnwidth]{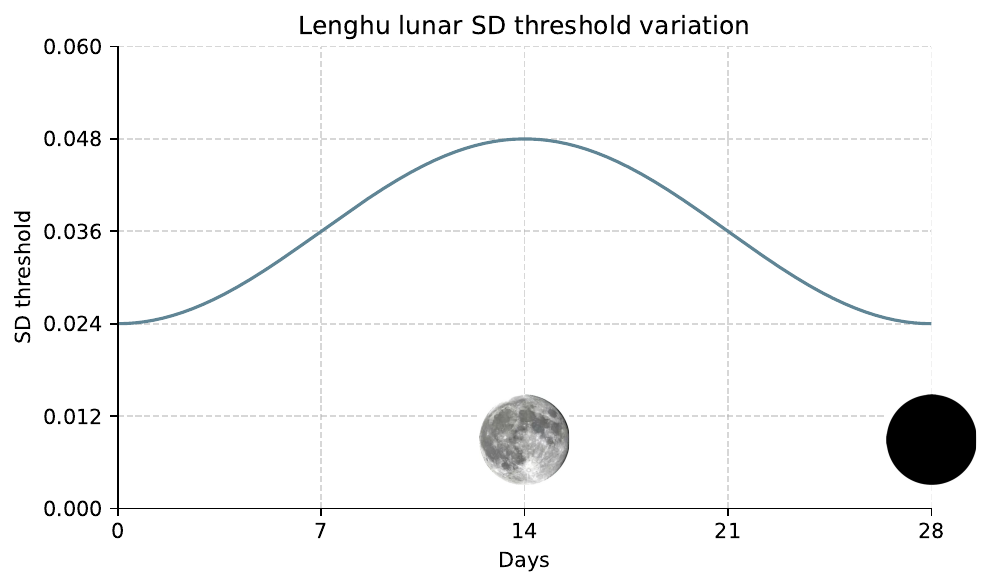}
    \caption{The standard deviation (SD) threshold $\sigma$ variation during a full lunation at Lenghu site. The SD threshold is described in mag arcsec$^{-2}$ with a 28-day moon cycle.}
    \label{threshold}
\end{figure}

\begin{figure*}
    \centering
	\includegraphics[width=0.9\textwidth]{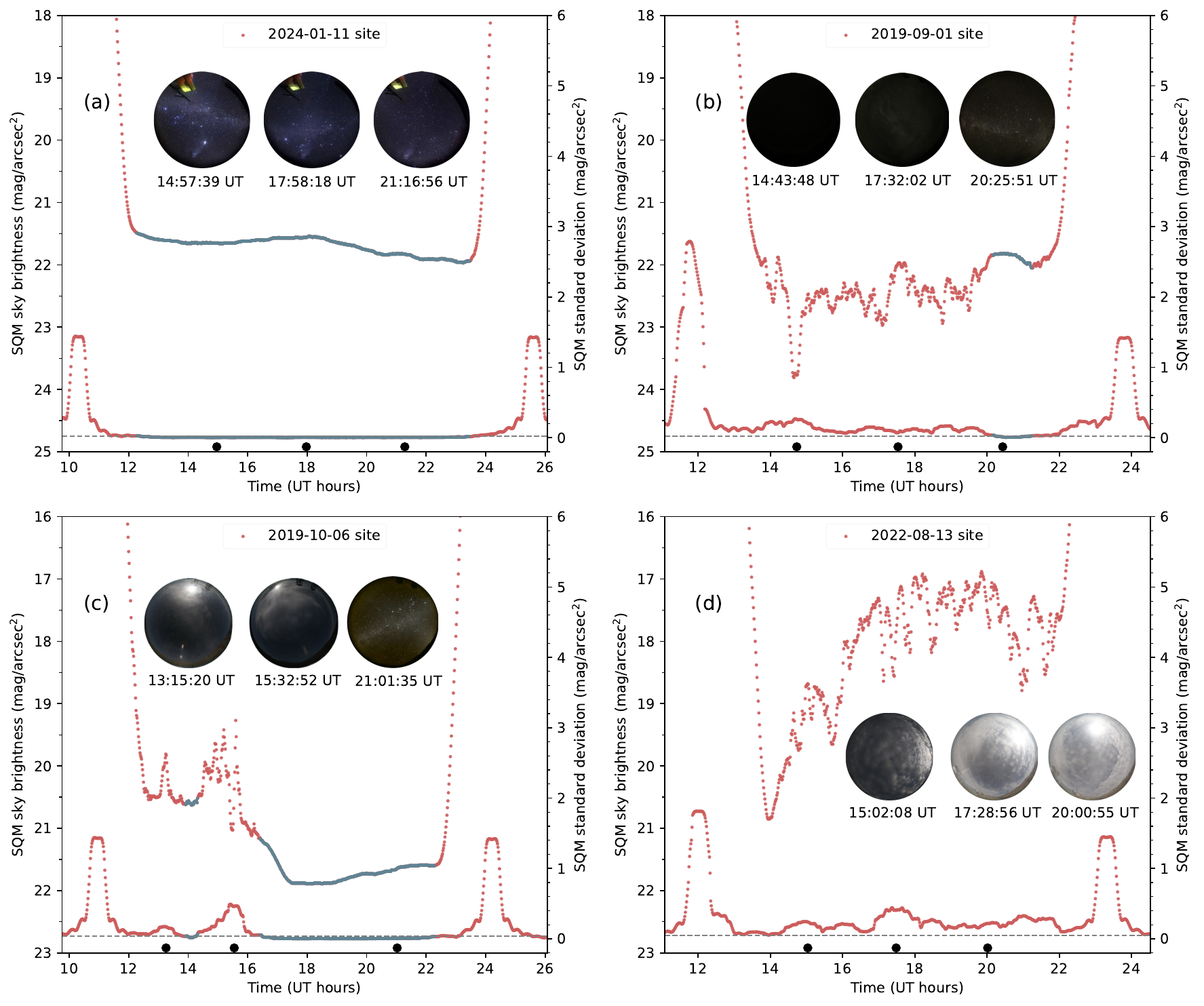}
    \caption{\textit{Panel a}: The SQM light curve of a fully clear night on 11 January 2024 is shown, blue segments denote "observable time", defined as "clear"; red segments are the periods that not suitable for observation, defined as "cloudy". The upper curve is the SQM light curve and the lower curve is the corresponding calculated standard deviation. The gray line of dashes marks the threshold $\sigma$=0.024 mag arcsec$^{-2}$ during a new moon. Three photos of the clear sky captured by LH-Cam are presented. \textit{Panel b}: The SQM light curve of 1st September 2019 around a new moon. The gray line of dashes marks the threshold $\sigma$=0.024 mag arcsec$^{-2}$. Before 20:00 UT, as shown in red segments, a cloudy sky is indicated, then the sky cleared for about an hour after 20:00 UT, which is illustrated in the blue segments. The calculation of the SQM empirical mathematical model perfectly matches the images captured by LH-Cam: the first two images show a dark night sky with stars obscured by clouds, while the last image shows that the clouds have dispersed, revealing a clear view of the Milky Way. \textit{Panel c}: The SQM light curve of 6 October 2019. The gray line of dashes marks the threshold $\sigma$=0.036 mag arcsec$^{-2}$ around a first quarter moon. Three representative images shot from LH-Cam which demonstrate ‘cloudy’, ‘passage of cirrus’ and ‘clear’ conditions are shown. \textit{Panel d}: The SQM light curve of a cloudy night on 13 August 2024. The dashed gray line marks the shreshold $\sigma$=0.048 mag arcsec$^{-2}$ around a full moon. Three photos of the cloudy sky captured by LH-Cam are presented, whose exposure were metered by SQM readings in realtime.}
    \label{fig:sqm}
\end{figure*}

For ground-based astronomical observations, the observable time (OT), dominated by the cloud cover, is a determinant among all site quality parameters. It is defined by the fraction of cloud-free nighttime, and second to which is the darkness of the night sky \citep{deng2021lenghu}. The sky darkness is greatly affected by light pollution, both natural and artificial \citep{wang2024phone}. Lenghu site suffers very little from artificial light pollution due to the fact that the only light source, the Lenghu town, is tiny in population and is 50 km away from the site. Therefore, the readout from the Sky Quality Meter (SQM) \citep{sqm2016} basically reflects the local natural sky darkness. Even though, in order to prevent future artificial light pollutions, the government of Qinghai Province has issued the Regulations for Lenghu Astronomical Observing Conditions Protection which came into effect on January 1, 2023. Artificial light sources will be strictly limited in the core zone centered at Lenghu site with a radius of 50 km, including the Lenghu Town. In the additional buffer zone with another 50 km extension of the radius, any artificial light sources point above horizon will be forbiden. The regulation prohibits any planning and building projects as well as activities that could impact astronomical observation in the core zone. The regulation also imposes restrictions on geological exploration, sightseeing, and other activities in this area to protect the astronomical observation environment. As China's first regulation for dark sky protection at professional astronomical sites, it serves as a reference for night sky protection at major international astronomical observatories.

To measure the darkness of the sky, we use the SQM that collects light from the entire sky from 400 nm to 600 nm and centers around the central wavelength of the Johnson V-band (500 nm) \citep{sqm2016}. SQM has a special optical design in front of the sensor so that the sensitivity is optimized towards its pointing. Normally, the device is put straight up the local zenith, the sensitivity rapidly decreases to less than 10 percent when the zenith angle exceeds 20$^\circ$. The readout of SQM is calibrated at the zenith and converted to sky brightness in mag arcsec$^{-2}$ in a passband larger than John V band. The sky brightness measured by a SQM during night changes very sensitively with the passage of clouds in the visual sky, leading to a very bumpy light curve. There are other factors that may cause light curve to fluctuate, for example, the rotation of the Earth leads to the bright part of the Milky Way passing by the view of SQM, which can result in slow and smooth changes of the sky brightness. The moonlight greatly brightens the night sky background, especially during a full moon. Atmospheric skyglow and zodiacal light also have some impact on the readout of the SQM, although both occur close to the local horizon and have little influence on SQM readings. 

Of course, the night sky brightness (NSB) measured by SQM is also critical for site testing, as it has been applied at many other sites (professional or amateur) worldwide \citep{sanchez2017sky, priyatikanto2023characterization}. We did not analyze it in this work because the site has been undergoing full-scale construction since 2022. The dust landed on the optical aperture of SQM during the day can change the NSB readings, but it still keeps the nightly variations of NSB a good assessment of cloud cover at nighttimes. It is worth mentioning that aerosols are the main light scattering factor in the cloudless sky and particulate matter concentration is crucial in NSB measuring \citep{scikezor2020impact, kocifaj2023systematic}. Aerosol optical depth (AOD) and related aerosol optics parameters or emission properties of the light sources also play non-negligible roles in the formation of skyglow \citep{wallner2023aerosol, kocifaj2020night, cavazzani2022launch}. Thus, the aforementioned parameters such as NSB and AOD will be analyzed in future site testing with fully constructed structure. The ageing effect of the SQM will also be taken into account in future analysis, as prolonged operation may eventually lead to noticeable threshold bias \citep{fiorentin2022instrument}. Extinction due to aerosol in the air is also one of the critical site parameters, which has not been monitored. This task cannot be put into practice at an early stage due to limited conditions on the mountain top. The earliest time when the mountain top was intact, SQM readings showed that the extreme NSB reaches 22.3 mag, which inferred that the extinction is minor compared to other sites. Also, due to constructions, we have to postpone the plan to when the situation becomes stable. This will also be a subject of further site characterization.

A tailor-made all-sky camera (LH-Cam) was also used to monitor the night sky clarity at Lenghu site, which is equipped with a commercial digital camera Canon EOS 750 and a fisheye lens (Sigma 4.5mm f/2.8). LH-Cam has taken an image of the whole sky every 20 minutes during the day and every 5 minutes during the night since March 2018. The exposure of the night-time images is regulated by SQM reading in real time to ensure image quality. LH-Cam work continuously at a duty cycle higher than 95\% in all times. By combining the LH-Cam and the SQM, we can obtain reliable OT of the site \citep{deng2021lenghu}. During clear nights, SQM readings form a smooth curve, varying only with the true starry sky background. When clouds sweep across the visible sky, the SQM output value fluctuates over time. The SQM is set to work at a sampling rate of every minute, so that cloudy conditions can be recorded by the light curve. We utilize the following technique to exclude the gradual change due to the presence of Moon or Milky Way, leaving the actual variation due to cloud and bad weather \citep{cavazzani2020sky}.

\begin{figure*}
    \centering
	\includegraphics[width=0.88\textwidth]{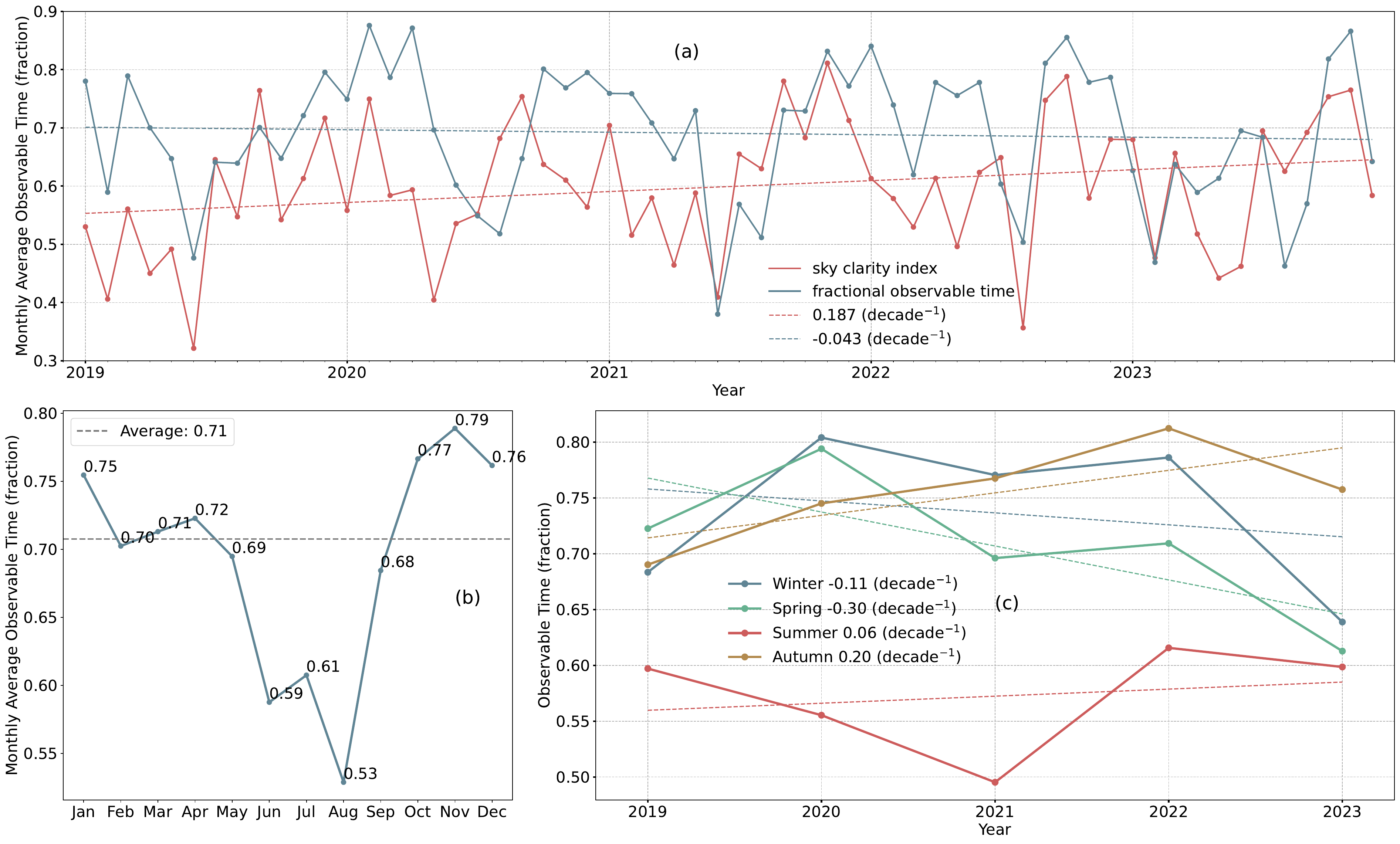}
    \caption{\textit{Panel a}: monthly average observable time fraction and sky clarity index calculated from ERA5 cloud cover reanalysis. \textit{Panel b}: monthly average observable time fraction obtained from superposed epoch analysis over 2019-2023. \textit{Panel c}: seasonal changes in observable time fraction at the Lenghu site.}
    \label{fig:observable}
\end{figure*}

\begin{table*}
	\centering
	\caption{Detailed observable time statistics in hours and days of Lenghu site. The annual SQM on-duty hours are shown in the second column, labeled as "Valid (h)". The next column labeled "Clear (h)" records the total time of clear nighttime sky in each year in hours, namely the OT. The fourth column "Clear/Valid" represents the proportion of clear-sky duration within the total valid nighttime working hour of SQM, which is the fractional OT plotted in Fig.~\ref{fig:observable}. The following "Bad/Valid" column is similar, representing the proportion of bad sky conditions compared to the valid working hours. The "Down/Total" column is the proportion of SQM downtime within the total nighttime duration, where the latter is calculated from astronomical dawn to the astronomical twilight next morning in hours, which is 3135.6 h for 2020 (leap year) and 3124.3 h for the rest. The subsequent three columns in the second row table enumerate the number of nights with more than 98\% / 50\% / 20\% of the nighttime that is observable. The last three columns tally the number of nights with contiguous fully clear time longer than 6 hours / 4 hours / 2 hours.}
	\label{tab:observable}
	\begin{tabular}{cccccccccccc} 
		\hline
		Year & Valid (h) & Clear (h) & Clear/Valid & Bad/Valid & Down/Total & 98\% clear & 50\% clear & 20\% clear & 6h clear & 4h clear & 2h clear\\
		\hline
		2019 & 3017.1 h & 2102.8 h & 69.70\% & 30.30\% & 3.43\% & 95 days & 247 days & 312 days & 126 days & 190 days & 267 days\\
		2020 & 2852.4 h & 2138.4 h & 74.97\% & 25.03\% & 9.03\% &103 days & 253 days & 310 days & 129 days & 203 days & 258 days\\
		2021 & 2949.5 h & 2072.2 h & 70.26\% & 29.74\% & 5.60\% &80 days & 242 days & 298 days & 127 days & 192 days & 260 days\\
        2022 & 2701.9 h & 2006.6 h & 74.27\% & 25.73\% & 13.52\% &76 days & 251 days & 294 days & 115 days & 194 days & 255 days\\
        2023 & 2469.3 h & 1608.0 h & 65.12\% & 34.88\% & 20.97\% &57 days & 187 days & 245 days & 90 days & 157 days & 215 days\\
		\hline
	\end{tabular}
\end{table*}

To use the SQM time series data (nightly light curves) to calculate the cloud cover and OT, we first apply a Butterworth lowpass filter to extract the low-frequency variations from the time series, which typically include slow changes such as planetary rotation and the natural variation of the night sky brightness over time and generally not associated with cloud passage or rapid weather changes. By subtracting this low-frequency signal from the original time series, we get the useful signals. To quantify their fluctuations, we compute the standard deviation at each time point using a window that spans 10 points (10 minutes) preceding and following, encompassing a total of 21 minutes.

We then use an empirical mathematical model to determine in which case the standard deviation value represents the passage of the cloud. A proper standard deviation threshold is the key to classify the clarity of night. The empirical correlation between the threshold and site magnitude can be described as \citep{cavazzani2020sky}:

\begin{equation}
    \sigma=-0.04545\times{M}+1.0500
	\label{eq:sigma}
\end{equation}

In which $\sigma$ represents the standard deviation threshold, and $M$ denotes the SQM reading in magnitudes.

At Lenghu site, on clear nights around the new moon, and with the bright part of the galactic disk far from the zenith, the background sky brightness is 22.3 mag arcsec$^{-2}$; while the moon is below the horizon, the background sky brightness is approximately 22.0 mag arcsec$^{-2}$ on average \citep{deng2021lenghu}. The standard deviation thresholds used in the eq.~(\ref{eq:sigma}) for 22.3 mag arcsec$^{-2}$ and 22.0 mag arcsec$^{-2}$ are 0.03647 mag arcsec$^{-2}$ and 0.05010 mag arcsec$^{-2}$, respectively. To ensure a safe detection of cloudy night sky time, we have selected a more stringent standard deviation threshold $\sigma$= 0.036 mag arcsec$^{-2}$.

Another complication for the sky background is due to the combination of cloud and the Moon. During a full moon, the fluctuations in light curve caused by clouds passing tend to be more significant, mainly due to illumination of moon light on clouds (in the same way as that of near big cities). Therefore, a higher threshold should be set. On the contrary, during a new moon, we set a lower threshold. We use a cosine function with a period equal to a lunar cycle to modulate the threshold, with a maximum value during the full moon and a minimum value during the new moon, so the threshold of a full lunation can be calculated as:

\begin{equation}
    \sigma(x) = \sigma - \frac{\sigma}{3} \cos\left(\frac{2 \pi}{T} (x)\right)
	\label{eq:sigma(x)}
\end{equation}

where \textit{x} is the number of days counting from the time of the last new moon, $\sigma$ is 0.036 arcsec$^{-2}$ as given before. The standard deviation threshold variation during a full lunation is shown in Fig.~\ref{threshold}: the threshold $\sigma$ is 0.024 arcsec$^{-2}$ during a new moon and 0.048 arcsec$^{-2}$ during a full moon.

In Fig.~\ref{fig:sqm}, we depicted examples of four days: two nights around a new moon, a night around a first quater moon and a night during a full moon. The SQM light curve and the corresponding calculated standard deviation are given, together with three representative images shot from LH-Cam for each night. From the comparison between the blue/ red segments calculated by the SQM empirical model and the photos captured by LH-Cam, we can see that the empirical model fits the reality well.

We compiled the monthly-averaged fraction of OT for the years 2019-2023 and used ERA5 reanalysis data to calculate the mean cloud cover for the Lenghu site. In Fig.~\ref{fig:observable}a, we present an overview of their monthly average variation trends. To facilitate a better comparison with OT, we subtract the mean cloud cover from 1 and denote it as the sky clarity index. The fitted blue line indicates that the changing rate of OT from 2019-2023 is -0.043 decade\(^{-1}\), showing a slight decrease over the last five years, while the corresponding trend of sky clarity index is 0.187 decade\(^{-1}\), exhibiting a relatively significant increasing trend.

Fig.~\ref{fig:observable}b displays the monthly average OT derived from superposed epoch analysis over five years. We obtain seasonal characteristics of the OT at Lenghu site: November has the highest average fraction of OT, 0.79, while August has the lowest, 0.53. Among them, the observable time fraction gradually decreases, with two minor local peaks in April and July, and then rises abruptly from August to November. The total average OT accounts for 0.71 among total nighttime duration (from astronomical dusk to astronomical dawn).

Fig.~\ref{fig:observable}c outlines the seasonal variation in fractional OT from 2019-2023. We apply meteorological season in this paper, with the average nighttime fractional OT for spring (March-May), summer  (June-August), autumn (September-November) and winter (December-February) being 0.71, 0.58, 0.75, 0.74. The fitting line over these five years shows the trends are -0.30 decade\(^{-1}\), 0.06 decade\(^{-1}\), 0.20 decade\(^{-1}\), -0.11 decade\(^{-1}\), respectively. Autumn boasts the best fraction of OT, while summer has the worst.

In Table~\ref{tab:observable}, we compile the detailed OT in hours and days. It is worth mentioning that due to the SQM's unrecorded time in 2023 exceeding 20\% of the total nighttime, it may not accurately reflect 2023's actual OT.

\subsection{Temperature}

\begin{figure*}
	\includegraphics[width=\textwidth]{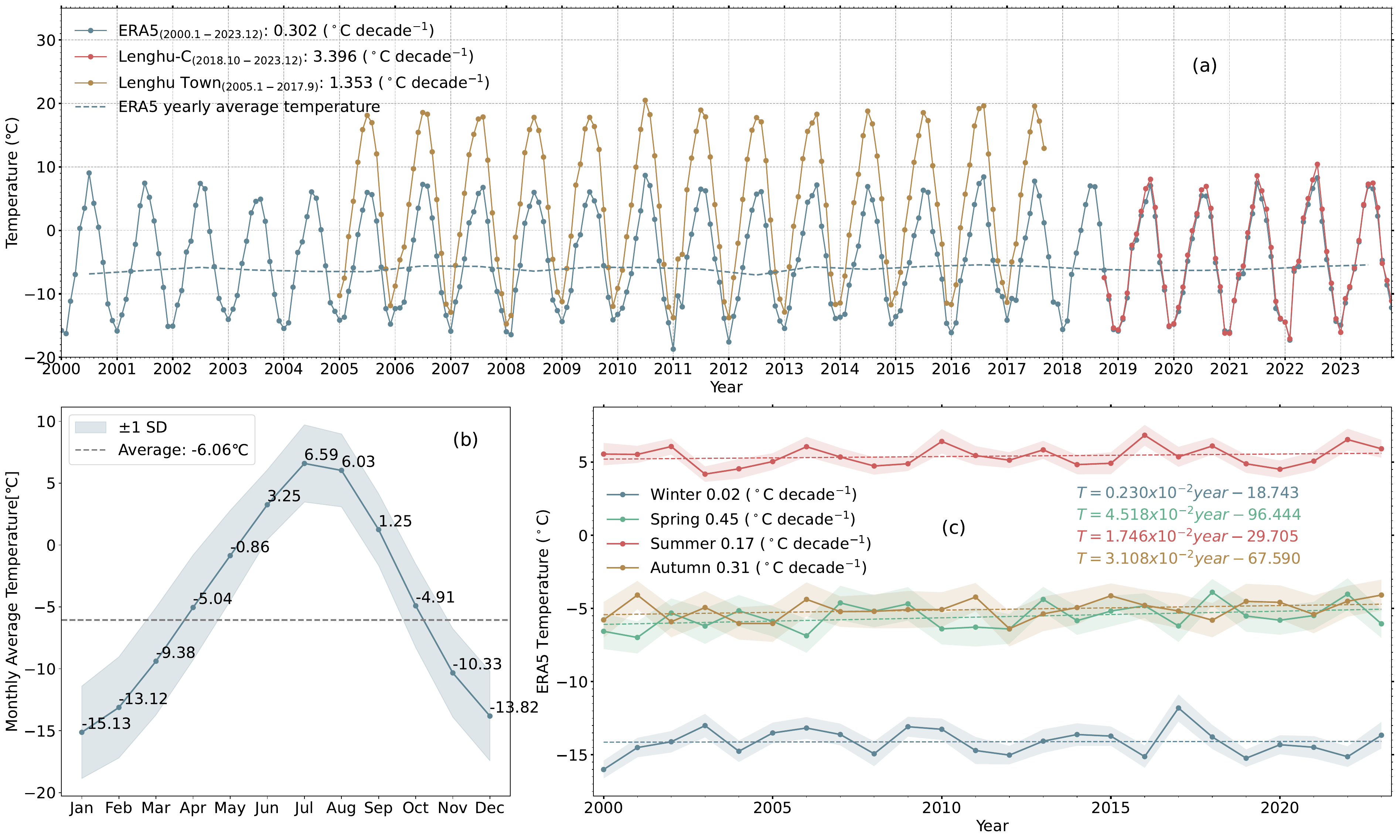}
    \caption{\textit{Panel a}: monthly average nighttime temperature measured from ERA5 reanalysis, weather station data and Lenghu Town data. A linear least squares regression is performed to calculate the ascending/decreasing trend per decade. The blue dotted line denotes the yearly average temperature obtained from ERA5 reanalysis.} \textit{Panel b}: monthly average nighttime temperature obtained from superposed epoch analysis over 2000-2023. \textit{Panal c}: seasonal variations in temperature obtained from ERA5 at Lenghu site. The shaded area shows the standard deviation for each season.
    \label{fig:temp}
\end{figure*}

\begin{figure}
    \centering
	\includegraphics[width=0.82\columnwidth]{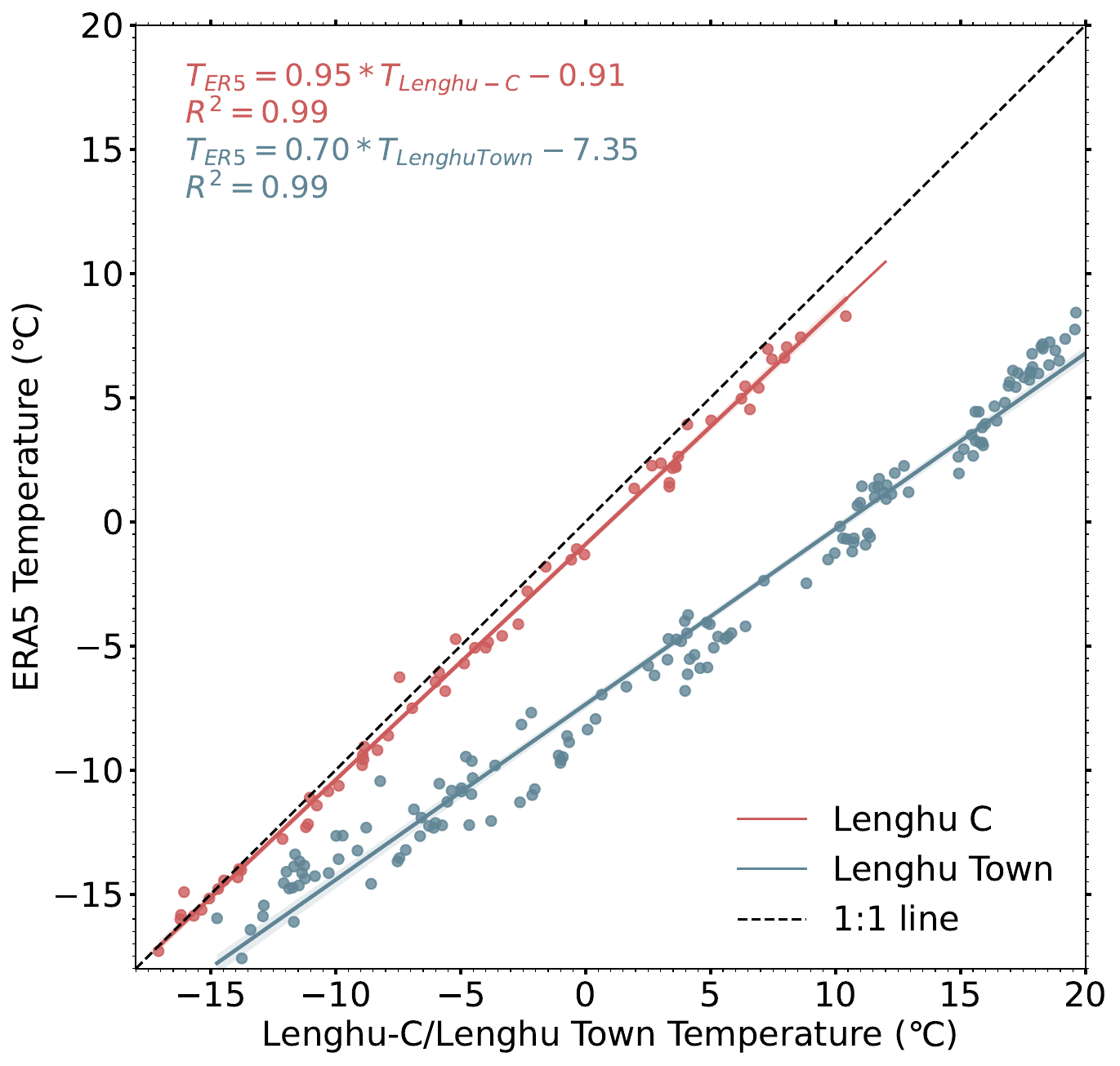}
    \caption{Red line marks the correlation between ERA5 data with weather station data in temperature from 2018.10 to 2023.12. Blue line marks the correlation between ERA5 data with Lenghu Town data in temperature from 2005.1 to 2017.9. The shaded areas shows the 95\% confidence intervals.}
    \label{fig:tempcorr}
\end{figure}

In order to make the application of the low spatial resolution ERA5 data to site quality analysis meaningful, we tested the consistency between the data from ERA5 and that of the local weather station at Lenghu-C. The comparison was made for the period between October 2018 and December 2023, the results are shown in Fig.~\ref{fig:tempcorr}. The correlation between ERA5 data and the ground station data is presented. The fitting shows excellent consistency, with an R$^{2}$ value of 0.99 or higher. The average systematic error between ERA5 and Lenghu-C is -0.50$^\circ$C. This consistency suggests that ERA5 can reliably be used to analyze the long-term trends at Lenghu site. Fig.~\ref{fig:tempcorr} also includes a comparison between the ERA5 temperature at the Lenghu-C site and the temperature in Lenghu Town. It is noted that, since the ERA5 data corresponds very well with the data from weather station at Lenghu-C and there was no measurements from 2005 to 2017 at Lenghu-C, the ERA5 data is used to compare with the meteorological data from Lenghu Town provided by the Haixi Meteorological Administration, as shown by the blue points in Fig.~\ref{fig:tempcorr}. Lenghu-C's average nighttime temperature is lower than that of Lenghu Town by 10.04$^\circ$C, and the fluctuation in average nighttime temperature is lower than that at Lenghu Town by 2.42$^\circ$C.

Fig.~\ref{fig:temp}a shows the ERA5 temperature trend at the Lenghu-C site from 2000 to 2023 (24 years), as well as the temperature trend of Lenghu Town from January 2005 to September 2017 and the trend of Lenghu-C ground weather station from October 2018 to December 2023. The fitting line reveals a modest ascending in both Lenghu-C ERA5 reanalysis and Lenghu Town data, with a rate of 0.302$^\circ$C and 1.353$^\circ$C per decade, respectively.

Fig.~\ref{fig:temp}b presents monthly average temperature derived from superposed epoch analysis over 2000-2023 using ERA5 reanalysis. It is shown that the highest nighttime temperature of 6.59$^\circ$C occur in July while the lowest nighttime temperature of -15.13$^\circ$C occur in January. The overall average nighttime temperature at Lenghu-C is -6.06$^\circ$C.

Fig.~\ref{fig:temp}c is the seasonal variation of nighttime temperatures in Lenghu-C from 2000 to 2023 using ERA5. The average nighttime temperature for spring, summer, autumn and winter are $-5.57$ $^\circ$C, 5.41$^\circ$C, -5.07$^\circ$C and -14.08$^\circ$C, respectively. Generally, except for the summer season, nighttime temperatures in the Lenghu area are all below 0$^\circ$C. The average nighttime temperature in winter remains constant, showing a negligible increase of 0.02$^\circ$C decade\(^{-1}\). Rising trends are observed in spring, summer, and autumn, with increasing rate of 0.45$^\circ$C decade\(^{-1}\), 0.17$^\circ$C decade\(^{-1}\), 0.31$^\circ$C decade\(^{-1}\), respectively. In addition, we have observed a pronounced 3-4 year cyclical variation in average winter nighttime temperatures, which may be associated with interannual oscillations of global circulation and oceanic modes.

\subsection{PWV}

\begin{figure*}
	\includegraphics[width=\textwidth]{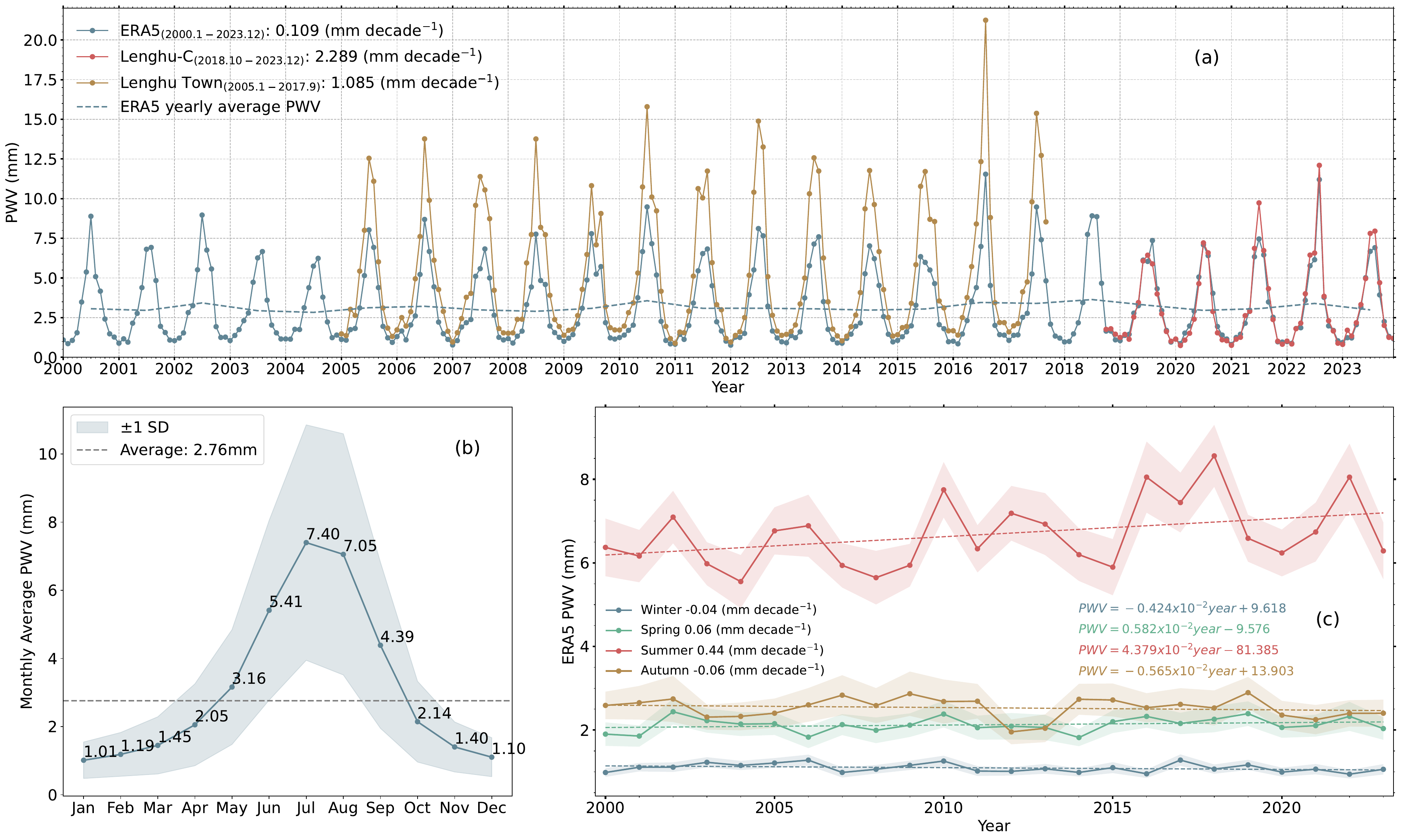}
    \caption{\textit{Panel a}: monthly average nighttime PWV measured from ERA5 reanalysis, weather station data and Lenghu Town data. A linear least squares regression is performed to calculate the ascending/decreasing trend per decade. The blue dotted line denotes the yearly average PWV obtained from ERA5 reanalysis. \textit{Panel b}: monthly average nighttime PWV obtained from superposed epoch analysis over 2000-2023. \textit{Panal c}: seasonal variations in PWV obtained from ERA5 at Lenghu site. The shaded area shows the standard deviation for each season.}
    \label{fig:pwv}
\end{figure*}

\begin{figure}
    \centering
	\includegraphics[width=0.82\columnwidth]{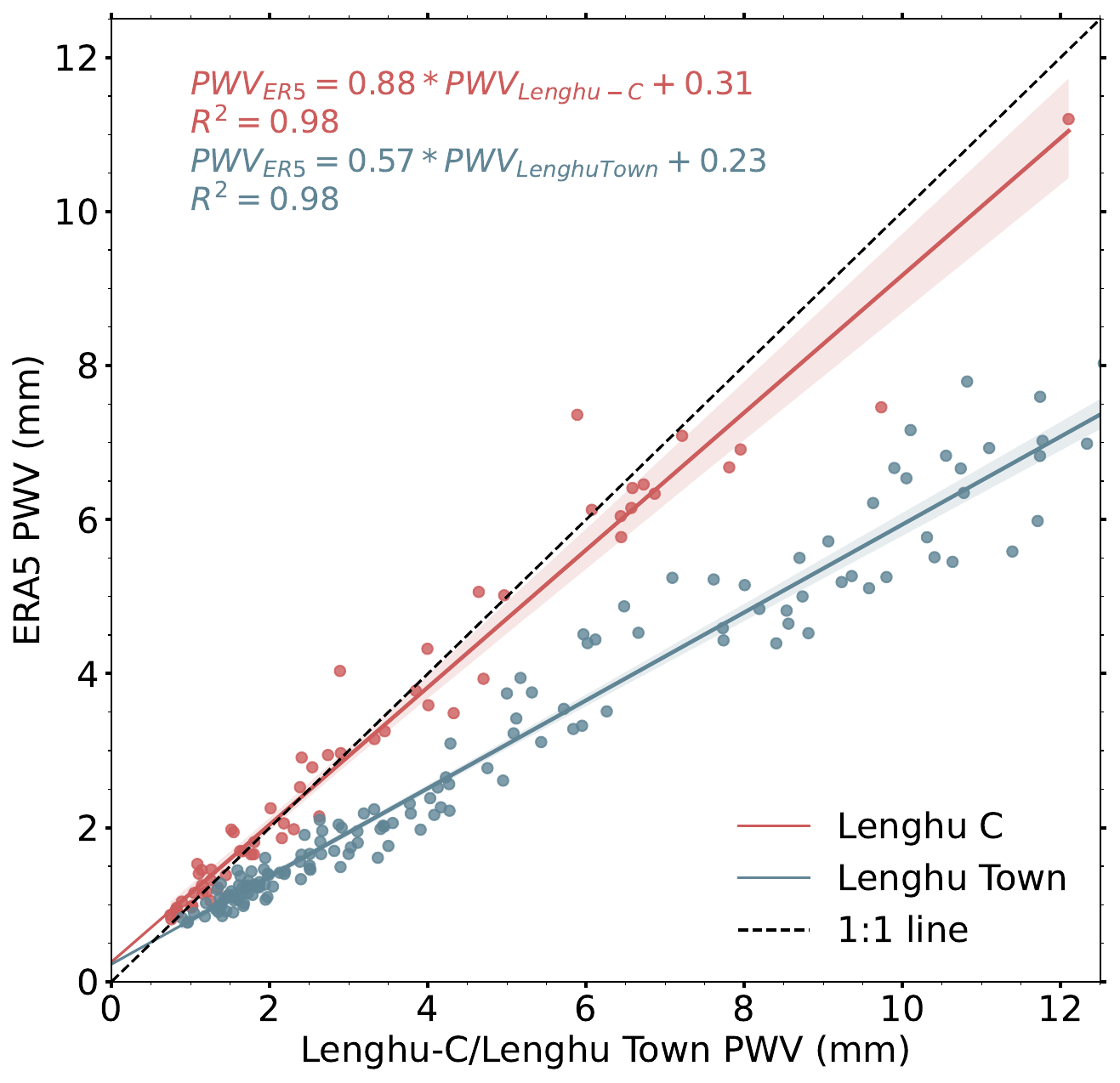}
    \caption{Same as Fig.~\ref{fig:tempcorr}, but for PWV.}
    \label{fig:pwvcorr}
\end{figure}

Precipitable water vapor (PWV) is the vertically integrated amount of water vapor from the surface to the top of the atmosphere \citep{peixoto1992physics}. To achieve cutting-edge scientific goals, observatories often conduct adaptive optics operating in infrared wavelengths (\citep{deng2021lenghu}) to which high PWV is fatal due to absorption. Not only is the light from celestial objects absorbed, but also the Earth's own thermal emissions can be backscattered into the telescope, both of them fluctuate over time due to atmospheric turbulence \citep{haslebacher2022impact}. Thus, a lower PWV is an important indicator of optical astronomical observatory sites quality. The Lenghu area, with its high altitude and arid climate, almost devoid of plants, also effectively avoids water vapor produced by plant transpiration. PWV can be calculated using the following formula \citep{qian2019empirical,wang2019comparison,deng2021lenghu,zhao2022long}:

\begin{equation}
    PWV = \frac{1}{\rho g} \int_{p_c}^{p_z} q \, dp
	\label{eq:PWV}
\end{equation}

where, $\rho$ is the density of liquid water, $g$ is the acceleration of gravity, $q$ is the specific humidity, $p_c$ and $p_z$ are the lower and upper limits of the pressure integral, respectively. The lower limit $p_c$ is the atmospheric pressure corresponding to the ground elevation of the Lenghu site and the upper limit $p_z$ is generally chosen to be the pressure at the tropopause as nearly all the water vapor and aerosols in the atmosphere are contained within the troposphere. The parameters $p_z$ and $q$ can be obtained from ERA5 reanalysis data. The integral's lower limit, $p_c$=607 hPa, is the average pressure at Lenghu-C. Since the ERA5 levels do not perfectly match the pressure levels at Lenghu-C, we have made corresponding corrections\citep{zhao2022long}.

We first made comparison between the three datasets for PWV analysis within the same time window as that for temperature above. As shown in Fig.~\ref{fig:pwvcorr}, the red line represents the correlation between the ERA5 data and the ground weather station's nighttime PWV data, with an R-squared value of 0.99, and a systematic error of -0.0275 mm, confirming that ERA5 reanalysis data can serve as a fine reference for long-term PWV trend analysis in the Lenghu area. The blue line represents the correlation of PWVs between Lenghu town based on local weather station and Lenghu-C calculated using ERA5. On average the former is  2.40 mm higher than the latter.

Fig.~\ref{fig:pwv}a provides an overview of the monthly variation trends of the aforementioned three datasets. The fitting line shows a slight upward trend of 0.109 mm decade\(^{-1}\) in Lenghu-C from 2000 to 2023 and 1.085 mm decade\(^{-1}\) in Lenghu Town from 2005.1 to 2017.9.

Fig.~\ref{fig:pwv}b displays the monthly average PWV obtained with the superposed epoch analysis from 2000-2023 using ERA5 reanalysis. It shows that the PWV at the Lenghu site exhibits clear seasonal characteristics. Similar to temperature, Lenghu-C experiences the lowest monthly average PWV in January with 1.01 mm and the highest in July with 7.40 mm. April to July develop a rapid rise in PWV, followed by a swift decline from August to October. The annual average nighttime PWV is 2.76 mm, with the majority of the PWV contribution concentrated summer, similar to the variation in temperature.

Fig.~\ref{fig:pwv}c shows the interannual variation of the average nighttime PWV for each season at Lenghu-C from 2000 to 2023. Summer has the highest PWV, followed by autumn, spring, and winter. The trend of summer PWV exhibits a relatively clear upward trend of 0.44 mm decade\(^{-1}\). The trend in spring, autumn and winter are relative stable, which are 0.06 mm decade\(^{-1}\), -0.06 mm decade\(^{-1}\) and -0.04 mm decade\(^{-1}\), respectively.

\subsection{Relative Humidity}

\begin{figure*}
	\includegraphics[width=\textwidth]{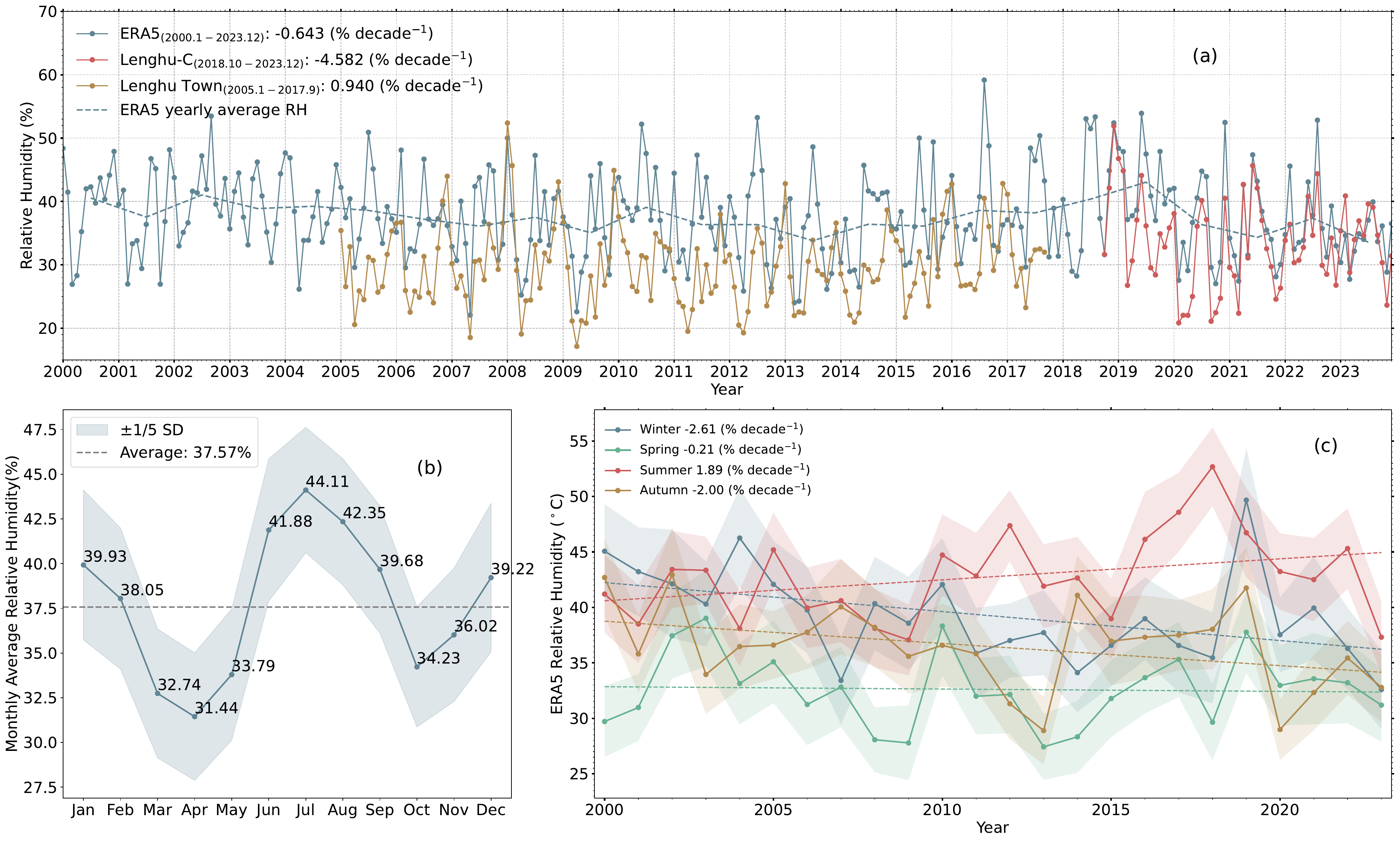}
    \caption{\textit{Panel a}: monthly average nighttime RH measured from ERA5 reanalysis, weather station data and Lenghu Town data. A linear least squares regression is performed to calculate the ascending/decreasing trend per decade. The blue dotted line denotes the yearly average RH obtained from ERA5 reanalysis. \textit{Panel b}: monthly average nighttime RH obtained from superposed epoch analysis over 2000-2023. \textit{Panal c}: seasonal variations in RH obtained from ERA5 at Lenghu site. The shaded area shows the standard deviation for each season.}
    \label{fig:rh}
\end{figure*}

\begin{figure}
    \centering
	\includegraphics[width=0.83\columnwidth]{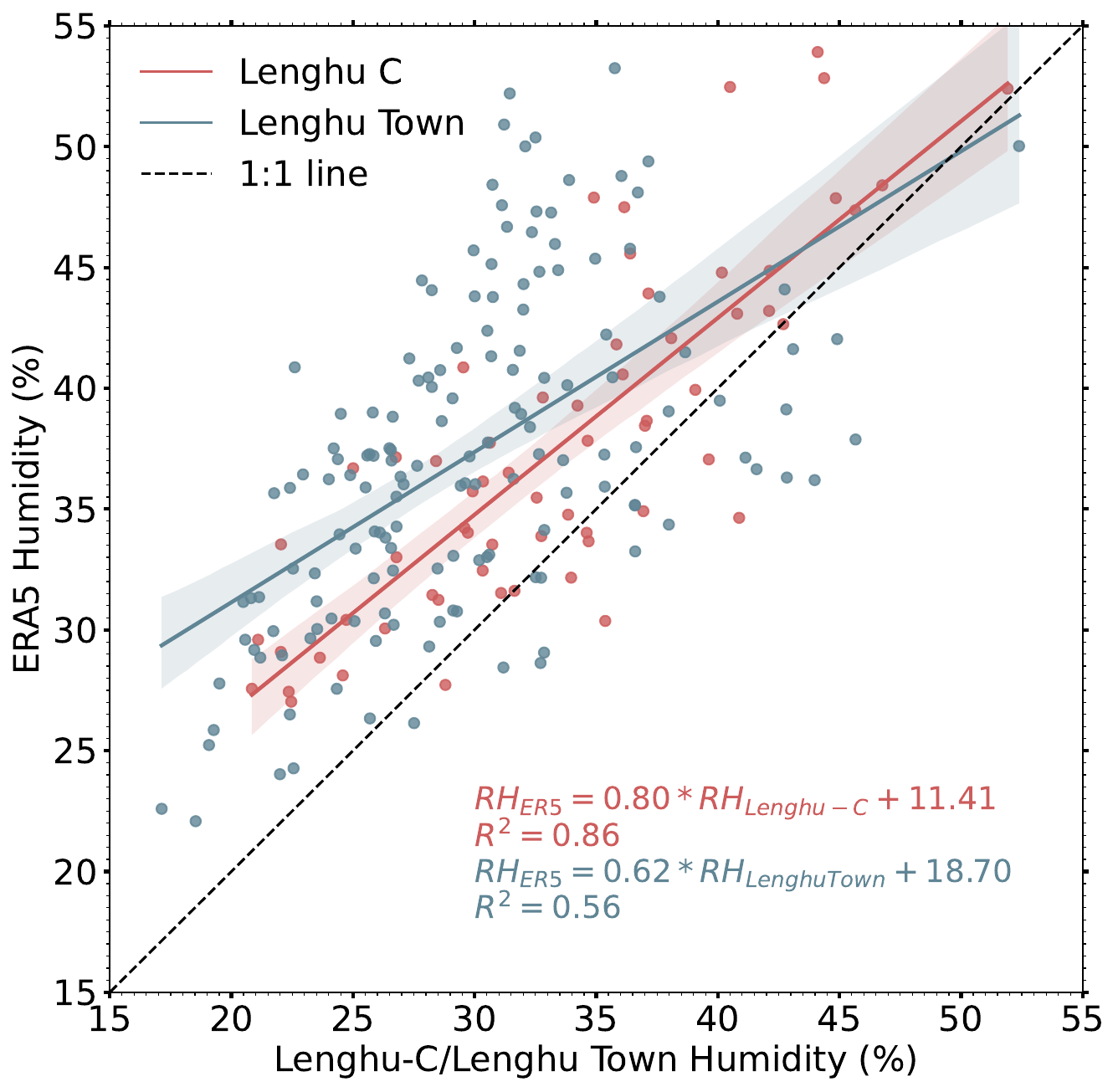}
    \caption{Same as Fig.~\ref{fig:tempcorr}, but for relative humidity.}
    \label{fig:rhcorr}
\end{figure}

The Relative Humidity (RH) at the surface level (measured by weather station on site) is an important restriction on telescope operations, high humidity often coexists with precipitation and cloud cover, as well as higher PWV. High RH increases the water vapor absorption in infrared spectrum, it also affects the size of aerosol particles which contribute to the scattering of light along the path. For astronomical observations, high RH conditions will cause frosting on the optical surfaces, and in long-term make oxidation, leaching, or even delamination of coatings \citep{bradley2006characterization}.

RH can be determined by the ratio of the actual specific humidity (q) to the saturation specific humidity (qs). This ratio is equivalent to the ratio of the water vapor pressure (e) to the saturated vapor pressure (es) \citep{peixoto1992physics}:

\begin{equation}
    RH=\frac{q}{qs}=\frac{e}{es}
	\label{eq:RH}
\end{equation}


Fig.~\ref{fig:rhcorr} shows the correlation between nighttime ERA5 RH data and ground-based meteorological station (Lenghu-C) RH data. The graph reveals an R-squared value of 0.86, indicating that the ERA5 reanalysis data can serve as a reference for the analysis of long-term RH trends in the Lenghu area. The systematic difference between ERA5 and Lenghu-C is 4.24\%.

Fig.~\ref{fig:rh}a displays the monthly mean RH trends from 2000 to 2023 derived from the ERA5 dataset (the top curve), the ground meteorological stations at Lenghu-C (the lower right curve) and at Lenghu Town (the lower left curve) respectively. The average nighttime RH at Lenghu-C (ERA5) is higher than Lenghu Town by 7.07\%, and the data from ERA5 indicates a trend of -0.643\% decade\(^{-1}\) from 2000 to 2023, maintaining a generally stable trend. The ground meteorological station at Lenghu-C shows a decreasing trend of -4.58\% decade\(^{-1}\) from 2019 to 2023. Lenghu Town experienced a slight increase in nighttime humidity from January 2005 to September 2017, which is 0.94 decade\(^{-1}\).

Fig.~\ref{fig:rh}b presents the variations of the monthly average RH for Lenghu-C from 2000 to 2023 using superposed epoch analysis. Lenghu's nighttime RH exhibits clear seasonal trends. The monthly average nighttime RH peaks twice a year, occurring in January and July, with averages of 39.93\% and 44.11\%, respectively. The two lowest values appear in April and October, with averages of 31.44\% and 34.23\%, respectively. The lowest and highest RH occurs in April and July, experiencing a rapid rise from the minimum to the maximum within three months. The second lowest and highest RH is in October and January, with each development from minimum to maximum or vice versa approximately in three months, demonstrating a biannual cyclical change within a year. 

Fig.~\ref{fig:rh}c shows the annual variation of nighttime RH in Lenghu-C from 2000 to 2023 using ERA5. The average RH for spring, summer, autumn and winter are 32.61\%, 42.77\%, 36.46\%, 39.15\% and the changing trends over these years are -0.21\% decade\(^{-1}\), 1.89\% decade\(^{-1}\), -2.00\% decade\(^{-1}\), -2.61\% decade\(^{-1}\), respectively, showing an obviously descending trend except the RH in summer.

\subsection{Wind Speed and Direction}

\begin{figure*}
	\includegraphics[width=\textwidth]{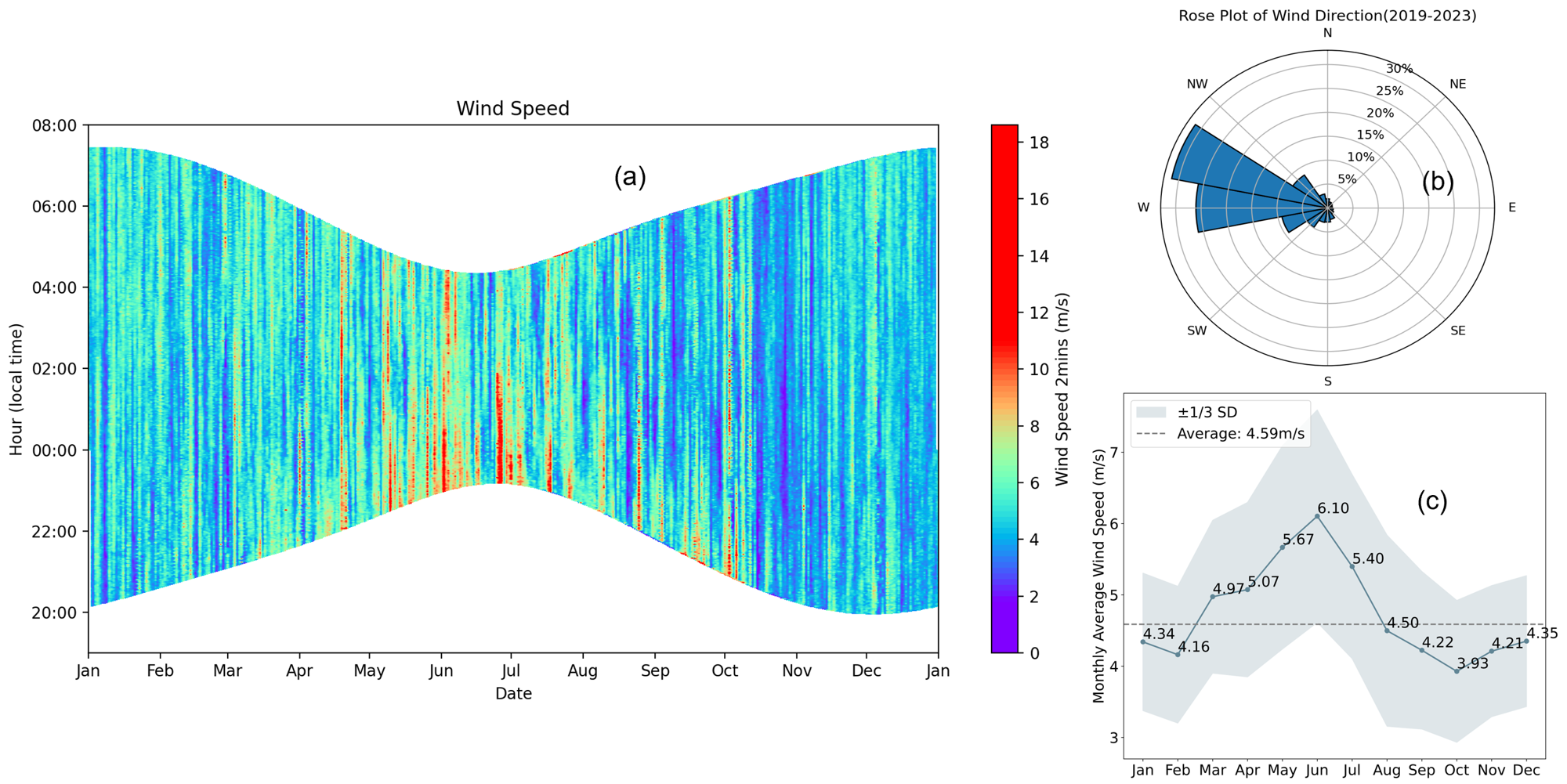}
    \caption{\textit{Panel a}: average nighttime surface wind speed measured from Lenghu-C weather station from 2019-2023. \textit{Panel b}: nighttime wind direction distribution obtained from ERA5 reanalysis over 2019-2023. \textit{Panal c}: monthly average nighttime wind speed obtained from superposed epoch analysis over 2019-2023.}
    \label{fig:wind}
\end{figure*}

\begin{figure*}
	\includegraphics[width=\textwidth]{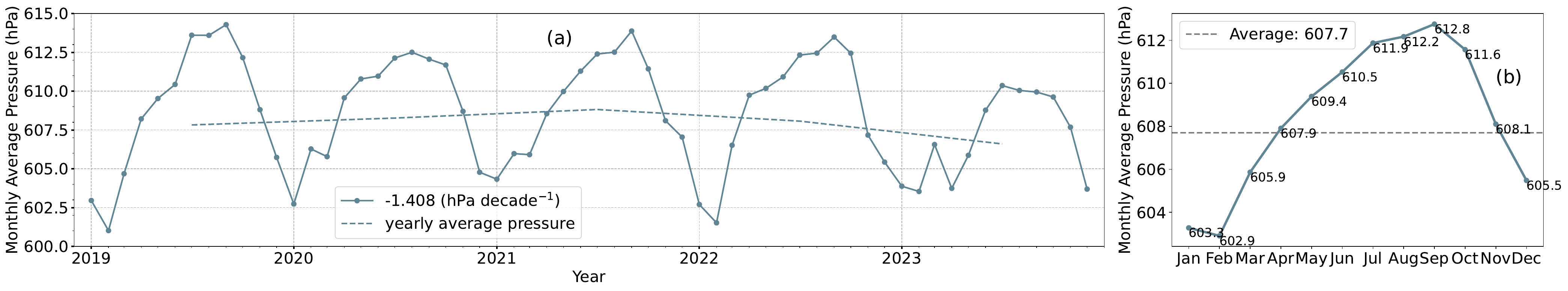}
    \caption{\textit{Panel a}: monthly average nighttime surface pressure measured from Lenghu-C weather station data over 2019-2023. \textit{Panel b}: monthly average nighttime pressure obtained from superposed epoch analysis.}
    \label{fig:pressure}
\end{figure*}

Surface wind has multifaceted impacts on astronomical observations. Excessively strong surface winds induce physical vibrate of the telescope, affecting pointing stability and tracking accuracy, thereby degrading the quality of the images \citep{bradley2006characterization}. Extremely strong wind speed may even damage exposed facilities. Additionally, strong winds may increase the density of aerosols and dust particles in the atmosphere which scatter and/or absorb celestial light especially in the infrared band. The dust deposits on the telescopes' optical and mechanical components, necessitating frequent cleaning and maintenance. For the Lenghu site which is located at the summit of Saishiteng Mountain with a 1500m elevation drop from the Yadan landform below, natural barriers prevent most of the dust from reaching the site. Wind speed is also a critical factor of local turbulence, since strong winds alter the atmosphere's refractive index and light propagation, thereby impact seeing quality \citep{zhu2023astronomical}. Moderate surface winds, however, can be beneficial: it can help disperse the heat layers near surface, reducing turbulence generated by ground heating, thus improving the stability of the near-ground atmosphere and seeing quality \citep{bradley2006characterization}. It may also prevent the accumulation of clouds, extending the OT. Therefore, astronomical sites are typically set at places with relatively low wind speeds and stable directions. Meanwhile, modern telescope structure designs take wind resistance into consideration and use adaptive optics technology to mitigate the impact of turbulence on observation quality.

Fig.~\ref{fig:wind} provides a detailed analysis using the ground-based meteorological station (Lenghu-C) data from 2019-2023. Fig.~\ref{fig:wind}a shows the average surface wind speed using superposed epoch analysis over these five years. The wind speed is higher around the dawn and dusk. Fig.~\ref{fig:wind}b displays the rose chart of wind direction, which is primarily concentrate in the direction of west by north. Fig.~\ref{fig:wind}c is the monthly average nighttime wind speed obtained from superposed epoch analysis. The average wind speed is 4.59m/s over these five years. It is clearly that the highest average nightime wind speed occurs in July, with a value of 6.10 m s\(^{-1}\), and the lowest average nighttime wind speed of 3.93 m s\(^{-1}\) happens in October.

\begin{figure*}
	\includegraphics[width=0.95\textwidth]{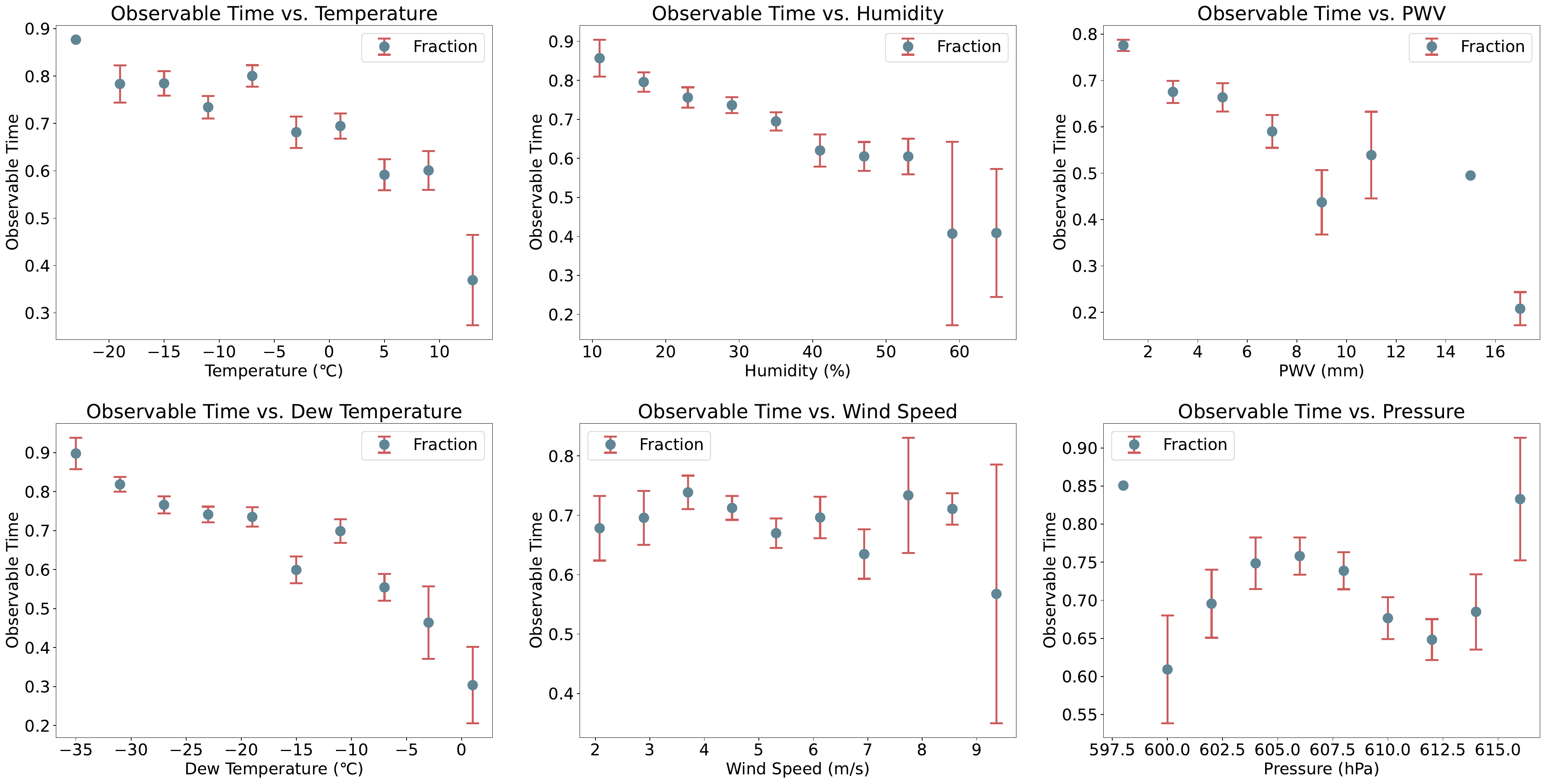}
    \centering
    \caption{Weekly average of observable time fraction against the corresponding weekly average of meteorological parameters using the data recorded from Lenghu-C ground station from 2019 to 2023. The errorbar presents the value of standard error of mean (sem) in each bin.}
    \label{fig:errorbar}
\end{figure*}

\begin{table*}
	\centering
	\caption{Student’s t-test for correlation between observable time fraction and meteorological parameters.}
	\label{tab:corr}
	\begin{tabular}{cccccc} 
		\hline
		parameter 1 & parameter 2 & PCC & P-value &\\
		\hline
		& Temperature & -0.378 & 1.55e-10 & correlated\\
        & Humidity & -0.373 & 2.55e-10 & correlated\\
        Observable time fraction & PWV & -0.520 & 5.78e-20 & correlated\\
        & Dew Temperature & -0.509 & 4.58e-19 & correlated\\
        & Wind Speed & -0.087 & 0.157 & no correlation\\
        & Pressure & -0.070 & 0.252 & no correlation\\
		\hline
	\end{tabular}
\end{table*}

\subsection{Pressure}

Variations in surface atmospheric pressure directly disturb the density stratification in atmosphere, thus affect the refractive index and the transparency of atmosphere along the light path. Fluctuations in surface pressure are often associated with shifts in other meteorological parameters such as temperature and humidity, which further impact the atmospheric optical properties. For high-precision astronomical instruments such as spectrometers and interferometers, adjustments may be necessary to calibrate for these pressure changes. 

Fig.~\ref{fig:pressure}a provides an overview of the monthly variation trends of surface pressure measured from Lenghu-C ground weather station. The fitting line shows a downward trend of -1.408 hPa decade\(^{-1}\) in Lenghu-C from 2019 to 2023.

Fig.~\ref{fig:pressure}b displays the monthly average pressure obtained with the superposed epoch analysis using the same dataset. A clear seasonal characteristic is shown. Lenghu-C experiences the lowest monthly average pressure in January with 603.3 hPa and the highest in September with 612.8 hPa. The annual average nighttime pressure is 607.9 hPa.

\section{Discussion}
\label{sec:discussion}

\subsection{Observable Time and Meteorology Parameters}
\label{sec:ot_mp}

In this section, we compare the weekly averages of OT with the corresponding weekly averages of meteorological parameters using the data recorded from Lenghu-C ground station from 2019 to 2023. To clearly show the variation trends, the meteorological parameters are binned into intervals of 4.0$^\circ$C for temperature, 10\% for RH, 2.0 mm for PWV, 4.0$^\circ$C for dew temperature, 1 m s\(^{-1}\) for wind speed, and 2 hPa for pressure, respectively, as shown in Fig.~\ref{fig:errorbar}. The error bars show the standard error of mean in the corresponding bins. We also calculated the linear correlation coefficients between OT and these meteorological parameters, as shown in Table~\ref{tab:corr}. We estimated the correlation between them with a Student’s t-test. The Pearson’s correlation coefficient (PCC) obtained from the analysis is presented along with the corresponding p-value. The significance level is set at 5\% (p-value) \citep{press1992numerical}, p-value less than 0.05 indicates that the two parameters are correlated. It was found that the OT of Lenghu-C is significantly negatively correlated to temperature, humidity, PWV, and dew temperature. 

\begin{figure*}
	\includegraphics[width=0.8\textwidth]{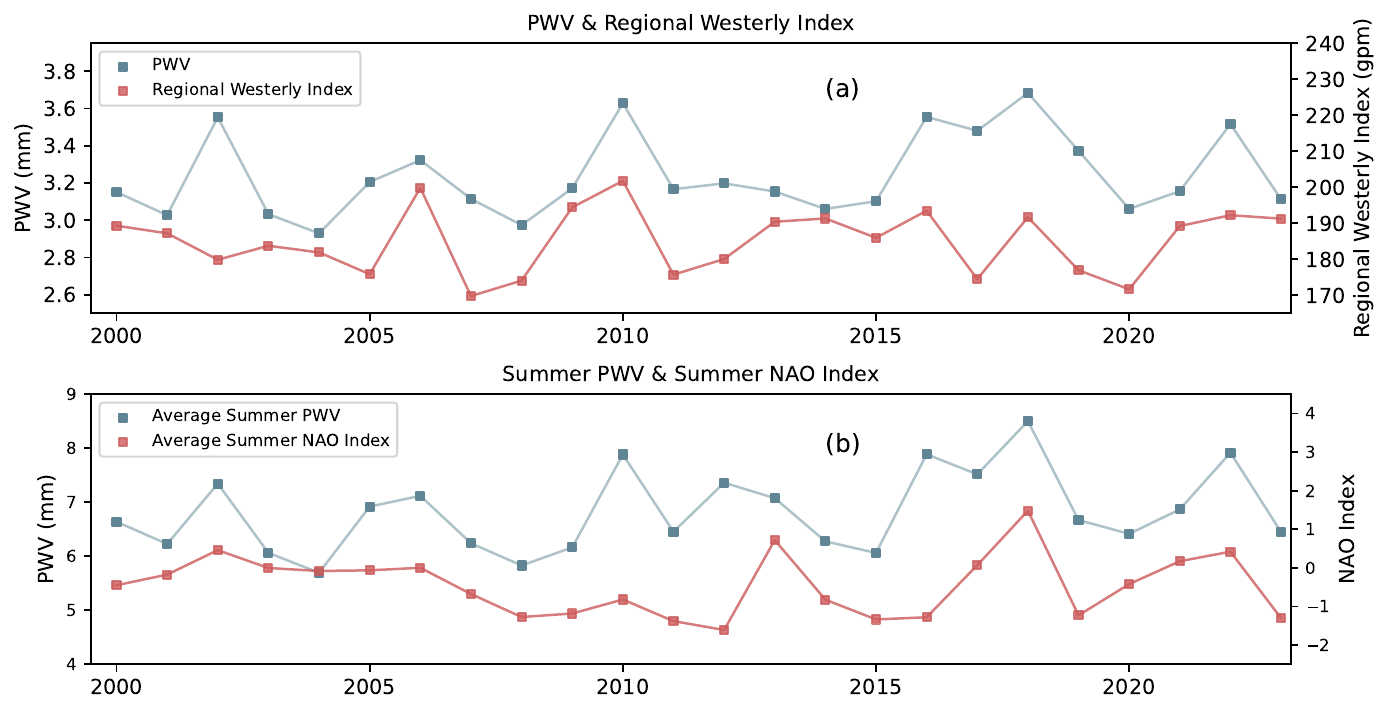}
    \centering
    \caption{\textit{Panel a}: variation of annual PWV and regional westerly index from 2000 to 2023. \textit{Panel b}: variation of average summer PWV and summer NAO index from 2000 to 2023.}
    \label{fig:nao}
\end{figure*}

\begin{table*}
	\centering
	\caption{Pearson Correlation Coefficient (PCC) and student’s t-test for the parameters in Fig.~\ref{fig:nao} (Lenghu annal PWV and regional weasterly index, average summer PWV and summer NAO index). The Spearman's rank correlation coefficient ($r_s$) and its P-value, indicating the strength and statistical significance of the monotonic relationship between the variables, are also given.}
	\label{tab:nao}
	\begin{tabular}{cccccc} 
		\hline
		parameter 1 & parameter 2 & PCC & P-value (PCC) & $r_s$ & P-value ($r_s$)\\
		\hline
		PWV & Regional Westerly Index & 0.341 & 0.103 & 0.356 & 0.088\\
        Summer PWV & Summer NAO Index & 0.422 & 0.040 & 0.356 & 0.088\\
        \hline
	\end{tabular}
\end{table*}

\subsection{NAO and Lenghu PWV}

Due to the specific location, Lenghu site holds a significantly important position in the climate system of the Northern Hemisphere. This region, mainly comprised of arid and semi-arid areas, is situated in the hinterlands of the Eurasian continent and the northern part of the Tibetan Plateau \citep{ding2023regional}. The variation in precipitation in Northwestern China is influenced by both internal and external circulations, with external circulation contributing up to 75\%, while internal circulation accounts for the remaining 25\% \citep{wu2019characteristics}. This indicates that external circulation plays a major role in atmospheric water vapor. The interdecadal variability is a non-negligible factor affecting precipitation among a large region surrounding the site. The ocean far beyond the vast Asia-Europa continent, as the most crucial external force factor, can modulate interdecadal climate changes through air-sea interaction \citep{yu2003spatial}.

\begin{figure*}
	\includegraphics[width=0.8\textwidth]{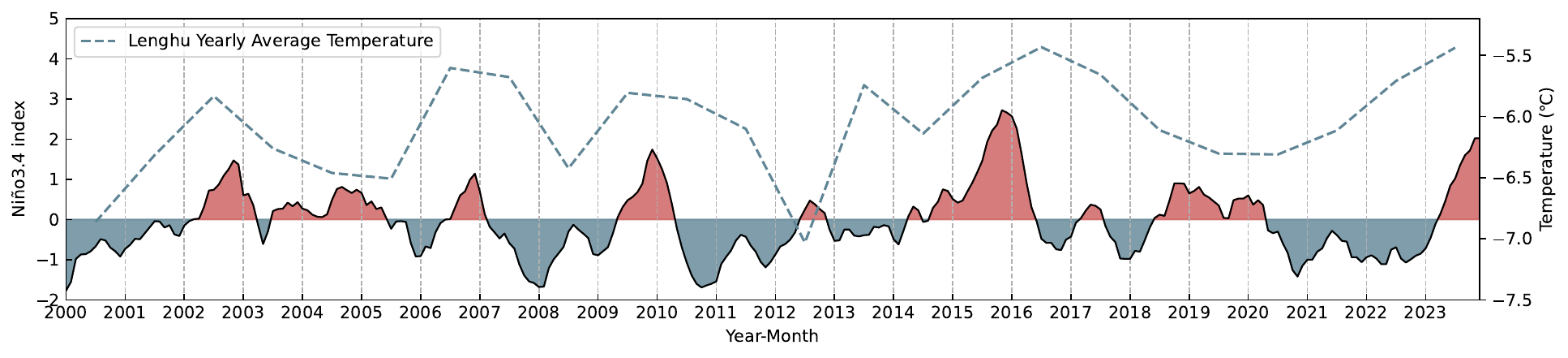}
    \centering
    \caption{Blue dotted line shows the trend of Lenghu yearly average temperature, the black curve shows the Niño3.4 index from 2000 to 2023.}
    \label{fig:t_enso}
\end{figure*}

\begin{figure*}
	\includegraphics[width=0.8\textwidth]{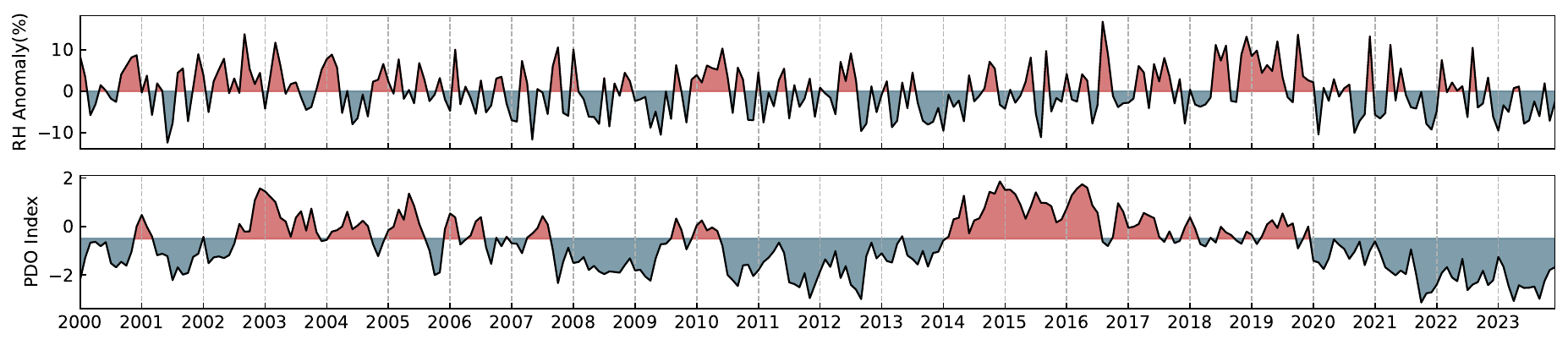}
    \centering
    \caption{The upper panel shows RH anomaly variation and the lower panel shows the the PDO index.}
    \label{fig:rh_pdo}
\end{figure*}

\begin{table*}
	\centering
	\caption{Student’s t-test for correlations between Lenghu RH and PDO/ENSO, The lags of month with highest Pearson Correlation Coefficient (PCC) is calculated by cross-correlate function.}
	\label{tab:rh}
	\begin{tabular}{cccccc} 
		\hline
		parameter 1 & parameter 2 & lags(month) & PCC & P-value &\\
		\hline
		RH & PDO & 0 & 0.129 & 0.0289 & correlated\\
        & ENSO & 0 & 0.072 &  0.222 & no-correlation\\
        \hline
        RH & PDO & 3 & 0.208 & 4.21e-05 & correlated\\
         & ENSO & 4 & 0.181 & 2.19e-04 & correlated\\
        \hline
	\end{tabular}
\end{table*}

The Westerlies and the Monsoon are the two major circulation systems that influence the climate in China. The domains and intensities of these two systems' influences vary with their forcing sources, such as global climate change, latitudinal and meridional thermal differences, and land-sea thermal contrasts \citep{ke2005atmospheric}.

The North Atlantic Oscillation (NAO) is a meteorological phenomenon characterized by fluctuations in the sea-level pressure (SLP) difference between the Icelandic Low and the Azores High over the North Atlantic. Through the oscillation in the intensity of the Icelandic Low and the Azores High, NAO governs the strength and direction of the westerlies at mid-latitudes \citep{NAO}. To a great extent, it exerts control over the dipole oscillation of summer precipitation and the distribution of moisture transport over the Tibetan Plateau by modulating the atmospheric circulation above and around, among which the westerlies is representative \citep{liu2015impact}.

To investigate whether the water vapor transport in the Lenghu region is also influenced by the westerlies as mentioned above, we calculated the westerly index, which effectively describes the strength of zonal westerlies in the temperate regions. It is traditionally definited by the difference in average geopotential height of 500 hPa across the latitude circle \citep{wen2004influences, wang1963preliminary}. Here, we apply the range of 70$^\circ$E to 110$^\circ$E and 35$^\circ$N to 50$^\circ$N to calculate the regional westerly index based on the vertical shear field of zonal wind \citep{li2008interrelationship}:

\begin{equation}
    I = \frac{1}{n} \left[ \sum_{\lambda=1}^{n} H(\lambda, 35N) - \sum_{\lambda=1}^{n} H(\lambda, 50N) \right]
	\label{eq:I}
\end{equation}

Here, H represents the averaged geopotential height field along the latitudes 35$^\circ$N and 50$^\circ$N, and $\lambda$ denotes the chosen longitudinal points, spaced at 2.5$^\circ$ intervals, beginning at 70$^\circ$E and end at 110$^\circ$E, which makes \( n = \left( \frac{110 - 70}{2.5} \right) + 1 = 17 \).

Fig.~\ref{fig:nao}a displays the trend of annual average PWV for Lenghu-C from 2000 to 2023 and the annual average regional westerly index, the result shows a quiet strong consistency in their trends. Except for 2002, both parameters have local peaks in 2006, 2010, 2016, 2018, and 2022, suggesting that the atmosphere moisture content in Lenghu region is related to wide-bound westerly transport within 70$^\circ$--110$^\circ$E, 35$^\circ$--50$^\circ$N. Fig.~\ref{fig:nao}b depicts the trend of summer average PWV and summer NAO index, the trend also shows consistency between them. Notably, the summer of 2010 and 2016 show significant PWV peaks without corresponding NAO index peaks, which coincide with strong El Niño events, indicating that the climate in Lenghu region is influenced not only by NAO, but also by multiple large-scale circulation and oceanic modes, which is more than complicated and this paper contains only a preliminary exploration of the correlations. The Pearson Correlation Coefficient (PCC) and the Spearman’s rank correlation coefficient ($r_s$), which indicates the significance of the monotonic relationship between the variables \citep{myers2013research}, as well as their P-values are given in Table~\ref{tab:nao}.

\subsection{ENSO and Temperature}

El Niño and the Southern Oscillation (ENSO) is the periodic fluctuation in the equatorial Pacific sea surface temperature (El Niño/La Niña) and the atmospheric pressure above it (Southern Oscillation). ENSO can have significant impact on global climate through air-sea interactions \citep{wang2000pacific, wu2017atmospheric(a), wu2017atmospheric(b), yeh2018enso, wang2019three}. The mid-latitude westerly wind system and the tropical monsoon circulation are the two major circulation systems that affect the Tibetan Plateau, with the former being one of the NAO's important outcomes as mentioned above, and the latter being dominated by ENSO. They interact with each other while acting on the Tibetan Plateau, this complex "westerlies-monsoon" interaction modulates the interannual climate variation of the Tibetan Plateau \citep{hu2021impact}. 

In Fig.~\ref{fig:t_enso} , we compiled the Lenghu annual average temperature and the ENSO (Niño 3.4) index from 2000 to 2023. We observed a significant increase in the annual average temperature at the Lenghu site during the occurrence of a strong El Niño event. In 2023, there was a notable rise in the overall annual average temperature, primarily contributed by the increase in temperatures during autumn and winter, while the temperatures in spring and summer decreased. As discussed in Sect.~\ref{sec:ot_mp}, there is a negative correlation between temperature and OT. Therefore, we suggest that the noticeable reduction in OT during the winter of 2023 was possibly associated with the effects of the strong El Niño event observed in that year.

\subsection{PDO and Relative Humidity}

The Pacific Decadal Oscillation (PDO) is a Pacific climate pattern often described as El Niño-like \citep{zhang1997enso}. Being parallel with the ENSO, the anomalies of PDO are classified into two phases, namely the warm phase and the cold phase, defined by the Northeast and tropical Pacific Oceans' temperature anomalies. During a warm/positive phase, the west Pacific is cooler and portion of the eastern ocean warms up; whereas during a cool/negative phase, a reverse pattern takes place \citep{zhang1997enso}.

In Fig.~\ref{fig:rh_pdo} , we present the monthly trend of RH anomaly and the PDO index from 2000 to 2023. The RH anomaly is calculated using the monthly average subtracted by 24 years' average of that month, which reveals a obvious positive correlation. Table~\ref{tab:rh} presents the correlation coefficients between RH anomaly and both the PDO index and the ENSO index. Additionally, cross-correlation analysis was conducted to explore the delayed response of RH to PDO/ENSO, revealing that the RH anomaly exhibits its highest correlation coefficient of 0.208 with the PDO index at a 3-month lag and a maximum correlation coefficient of 0.181 with the ENSO index at a 4-month lag. These facts in turn tell us that the RH in Lenghu region is modulated by large-scale oceanic modes, such as PDO and ENSO.

\section{Summary and conclusions}

\begin{figure*}
	\includegraphics[width=0.6\textwidth]{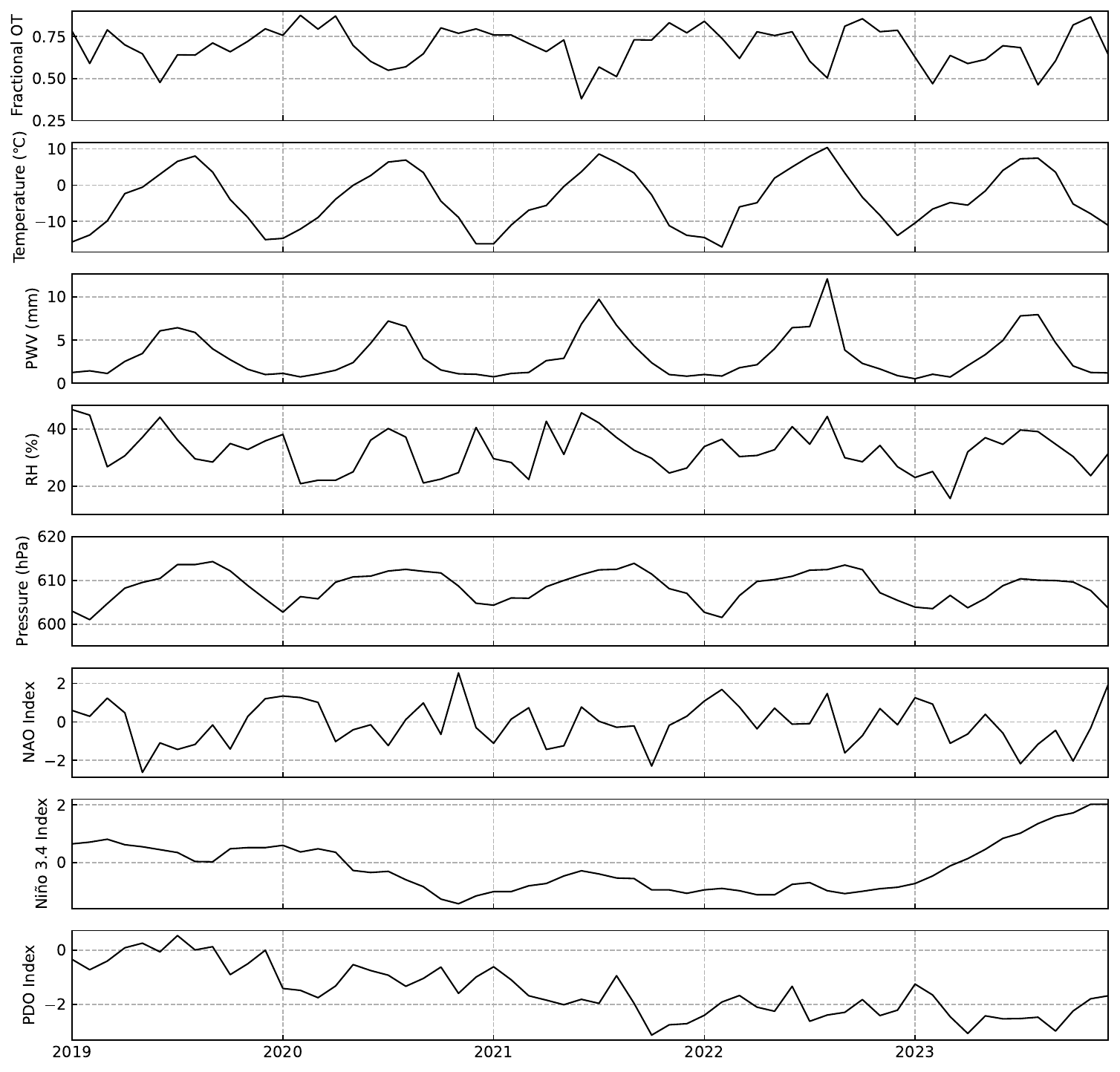}
    \centering
    \caption{The first panel shows the monthly fractional OT. The next four panels present the trend of monthly average temperature, PWV, RH and pressure measured from Lenghu-C ground weather station. The last three panels are the NAO index, Niño 3.4 index and the PDO index from 2019 to 2023.}
    \label{fig:sumfig}
\end{figure*}

In this paper, we provide a comprehensive analysis of the cloud cover in terms of OT of Lenghu observing site and its correlation with large-scale and long-term meteorological parameters. Fig.~\ref{fig:sumfig} is a summary figure that shows major collected data from 2019-2023. All the analysis are based on the local weather stations, long-term ERA5 reanalysis data and the historical weather records. The results can be summarized in the following:

\begin{enumerate}
      \item For the meteorological parameters such as surface air temperature, PWV, wind speed/direction, RH and surface pressure, the nighttime averages are respectively 6.06$^\circ$C (temperature), 2.76mm (PWV), 37.57\%(RH), 4.59m s\(^{-1}\)(wind speed) and 607.9hPa (pressure). A comparison was made between the site at the altitude of about 4200m (summit C) and the Lenghu town located at altitude of 2700m, with an extra altitue of about 1500m the temperature and PWV at the site is much lower than that of the town by 10.04$^\circ$C and 2.40mm, while RH is higher by 7.07\%.
      \item The fractional clear time (OT ratios) in 2019-2023 period are 69.70\%, 74.97\%, 70.26\%, 74.27\%, 65.12\%, respectively. Although a large fluctuation is seen in the fractional OT, the sky clarity index is experiencing an obvious ascent of 18.7\% decade\(^{-1}\). The weekly averages of fractional OT was found to be obviously negatively correlated with surface air temperature, RH, PWV, and dew temperature.
      \item The annual variability of meteorological parameters can be attributed in part to the influence of large-scale, or even global climate oscillations. Large-scale atmospheric circulation and air-sea interactions certainly have impacts on the meteorological conditions at the site, which determine long-term site quality variation and the scientific potential of all the facilities to be built. The westerly wind dominates the wind directions in Lenghu region and has great influence on water vapor transportation. We found a rather strong correlation between Lenghu PWV and westerly wind index, as well as the summer PWV with summer North Atlantic Oscillation (NAO) index, where the NAO is likely to govern the westerly circulation at mid-latitudes. During a strong El Niño event, an obvious increase is observed in the annual average temperatures. A correlation between Lenghu RH anomaly and PDO index has also been found, therefore we tend to conclude that RH in Lenghu region is also influenced by large-scale oceanic modes like PDO and ENSO.
      \item In a long period between 2000 and 2023, the temperature at the Lenghu site is stable in winter while showing a rising trend in spring, with the overall trend being slightly upwards. In terms of PWV, there is a slight decline in autumn and winter while has a significant increase in summer, also showing an overall rising. These observations suggest that the impact of global warming on the site are predominantly manifested during spring and summer. The rather good OT statistics is mainly due to the longer night-time and prevail climate conditions in autumn and winter. Thus we can further conclude that the Lenghu site suffers less  from global warming than most of the continental observatories.
      \item In 2023, a sizable decrease of OT at Lenghu site is recorded. Among all other factors leading to the reduction of OT, the frequent power outages at the site due to construction need to be considered, which caused interruption of all monitors including  SQM using which OT was calculated. The corresponding downtime exceeded 20\% of the night-time hours, and very unfortunately most of the downs occurred during autumn and winter of 2023. Additionally, large-scale climate modes also played a part. The decrease in OT during autumn and winter of 2023 stemmed from temperature increases, which can be attributed to the strong El Niño event in the year. Conversely, a negative phase of NAO during the summer reduced westerly wind moisture transport, leading to lower PWV and thus less pronounced reductions in summer OT. The winter temperature in 2023 has reached to a peak in 3-4 year cyclical variation. By the end of May, 2024, the fractional OT has reached 73.42\%.
   \end{enumerate}

The effects of global climate change on astronomical climate parameters can determine the scientific potential of the site, therefore further investigations are needed. Additionally, other crucial site parameters such as NSB, AOD and extinction are not currently fully considered due to site construction and will be studied systematically in the future. Understand the complex interplay between local meteorological conditions and global climate patterns is crucial for optimizing the long-term scientific operations of the Lenghu site and observatories worldwide.

\section*{Acknowledgements}

This work is mainly supported by the grant 2023FY101100 of the Ministry of Science and Technology of China (MOST) and the key project 12233009 of the National Natural Science Foundation of China (NSFC). The authors acknowledge NSFC for the support through grants 12273064 (FY \& LD), 42222408 (FH), 12322306 (XC). We are grateful to the long-term support and collaboration from Haixi Autonomous Prefecture Meteorological Bureau.

\section*{Data Availability}

Daily data for the Lenghu site, encompassing weather conditions, seeing, turbulence and more, are updated daily on Lenghu website \url{http://lenghu.china-vo.org/}. The ERA5 data is freely available from the Copernicus Climate Change Service Climate Data Store (CDS) at \url{https://cds.climate.copernicus.eu/cdsapp}. The global climate monitoring data like ENSO index, PDO index, and NAO index can be accessed from the website of National Oceanic and Atmospheric Administration: National Centers for Environmental Information at \url{https://www.ncei.noaa.gov/access/monitoring/products/}.
 



\bibliographystyle{mnras}
\bibliography{LRY_P1} 

\begin{thebibliography}{}
\makeatletter
\relax
\def\mn@urlcharsother{\let\do\@makeother \do\$\do\&\do\#\do\^\do\_\do\%\do\~}
\def\mn@doi{\begingroup\mn@urlcharsother \@ifnextchar [ {\mn@doi@} {\mn@doi@[]}}
\def\mn@doi@[#1]#2{\def\@tempa{#1}\ifx\@tempa\@empty \href {http://dx.doi.org/#2} {doi:#2}\else \href {http://dx.doi.org/#2} {#1}\fi \endgroup}
\def\mn@eprint#1#2{\mn@eprint@#1:#2::\@nil}
\def\mn@eprint@arXiv#1{\href {http://arxiv.org/abs/#1} {{\tt arXiv:#1}}}
\def\mn@eprint@dblp#1{\href {http://dblp.uni-trier.de/rec/bibtex/#1.xml} {dblp:#1}}
\def\mn@eprint@#1:#2:#3:#4\@nil{\def\@tempa {#1}\def\@tempb {#2}\def\@tempc {#3}\ifx \@tempc \@empty \let \@tempc \@tempb \let \@tempb \@tempa \fi \ifx \@tempb \@empty \def\@tempb {arXiv}\fi \@ifundefined {mn@eprint@\@tempb}{\@tempb:\@tempc}{\expandafter \expandafter \csname mn@eprint@\@tempb\endcsname \expandafter{\@tempc}}}

\bibitem[\protect\citeauthoryear{Bradley, Roberts, Bradford, Skinner, Nahrstedt, Waterson  \& Kuhn}{Bradley et~al.}{2006}]{bradley2006characterization}
Bradley E.~S.,  Roberts L.~C.,  Bradford L.~W.,  Skinner M.~A.,  Nahrstedt D.~A.,  Waterson M.~F.,   Kuhn J.~R.,  2006, Publications of the Astronomical Society of the Pacific, 118, 172

\bibitem[\protect\citeauthoryear{Bustos, Rubio, Ot{\'a}rola  \& Nagar}{Bustos et~al.}{2014}]{bustos2014parque}
Bustos R.,  Rubio M.,  Ot{\'a}rola A.,   Nagar N.,  2014, Publications of the Astronomical Society of the Pacific, 126, 1126

\bibitem[\protect\citeauthoryear{Cavazzani, Ortolani, Bertolo, Binotto, Fiorentin, Carraro, Saviane  \& Zitelli}{Cavazzani et~al.}{2020}]{cavazzani2020sky}
Cavazzani S.,  Ortolani S.,  Bertolo A.,  Binotto R.,  Fiorentin P.,  Carraro G.,  Saviane I.,   Zitelli V.,  2020, Monthly Notices of the Royal Astronomical Society, 493, 2463

\bibitem[\protect\citeauthoryear{Cavazzani et~al.,}{Cavazzani et~al.}{2022}]{cavazzani2022launch}
Cavazzani S.,  et~al., 2022, Monthly Notices of the Royal Astronomical Society, 517, 4220

\bibitem[\protect\citeauthoryear{Deng et~al.,}{Deng et~al.}{2021}]{deng2021lenghu}
Deng L.,  et~al., 2021, Nature, 596, 353

\bibitem[\protect\citeauthoryear{Fiorentin, Cavazzani, Ortolani, Bertolo  \& Binotto}{Fiorentin et~al.}{2022}]{fiorentin2022instrument}
Fiorentin P.,  Cavazzani S.,  Ortolani S.,  Bertolo A.,   Binotto R.,  2022, Measurement, 191, 110823

\bibitem[\protect\citeauthoryear{Haslebacher, Demory, Demory, Sarazin  \& Vidale}{Haslebacher et~al.}{2022}]{haslebacher2022impact}
Haslebacher C.,  Demory M.-E.,  Demory B.-O.,  Sarazin M.,   Vidale P.~L.,  2022, Astronomy \& Astrophysics, 665, A149

\bibitem[\protect\citeauthoryear{{Hersbach} et~al.,}{{Hersbach} et~al.}{2020}]{Hersbach2020}
{Hersbach} H.,  et~al., 2020, Quarterly Journal of the Royal Meteorological Society, 146, 1999

\bibitem[\protect\citeauthoryear{Hu, Zhou  \& Wu}{Hu et~al.}{2021}]{hu2021impact}
Hu S.,  Zhou T.,   Wu B.,  2021, Journal of Climate, 34, 3385

\bibitem[\protect\citeauthoryear{Hurrell, Kushnir, Ottersen  \& Visbeck}{Hurrell et~al.}{2003}]{NAO}
Hurrell J.~W.,  Kushnir Y.,  Ottersen G.,   Visbeck M.,  2003, The North Atlantic Oscillation: Climatic Significance and Environmental Impact.
American Geophysical Union

\bibitem[\protect\citeauthoryear{Ke-li, Hao  \& Hong-yan}{Ke-li et~al.}{2005}]{ke2005atmospheric}
Ke-li W.,  Hao J.,   Hong-yan Z.,  2005, Advances in Water Science, 16, 432

\bibitem[\protect\citeauthoryear{Kocifaj \& Bar{\'a}}{Kocifaj \& Bar{\'a}}{2020}]{kocifaj2020night}
Kocifaj M.,  Bar{\'a} S.,  2020, Monthly Notices of the Royal Astronomical Society: Letters, 500, L47

\bibitem[\protect\citeauthoryear{Kocifaj, K{\'o}mar, Lamphar, Barentine, Wallner  \& S}{Kocifaj et~al.}{2023}]{kocifaj2023systematic}
Kocifaj M.,  K{\'o}mar L.,  Lamphar H.,  Barentine J.,  Wallner  S 2023, Nature Astronomy, 7, 269

\bibitem[\protect\citeauthoryear{Li, Wang, Fu  \& Jiang}{Li et~al.}{2008}]{li2008interrelationship}
Li W.,  Wang K.,  Fu S.,   Jiang H.,  2008, Journal of Glaciology and Geocryology, 30, 28

\bibitem[\protect\citeauthoryear{Liu, Duan, Li, Shi, Yang, Zhang  \& Sun}{Liu et~al.}{2015}]{liu2015impact}
Liu H.,  Duan K.,  Li M.,  Shi P.,  Yang J.,  Zhang X.,   Sun J.,  2015, International Journal of Climatology, 35, 4539

\bibitem[\protect\citeauthoryear{Myers, Well  \& Lorch~Jr}{Myers et~al.}{2013}]{myers2013research}
Myers J.~L.,  Well A.~D.,   Lorch~Jr R.~F.,  2013, Research design and statistical analysis.
Routledge

\bibitem[\protect\citeauthoryear{Peixoto, Oort, Covey  \& Taylor}{Peixoto et~al.}{1992}]{peixoto1992physics}
Peixoto J.~P.,  Oort A.~H.,  Covey C.,   Taylor K.,  1992, Physics Today, pp 67--67

\bibitem[\protect\citeauthoryear{Press, Teukolsky, Vetterling  \& Flannery}{Press et~al.}{1992}]{press1992numerical}
Press W.~H.,  Teukolsky S.~A.,  Vetterling W.~T.,   Flannery B.~P.,  1992, Numerical recipes in C.
Cambridge university press New York, NY

\bibitem[\protect\citeauthoryear{Priyatikanto, Mumpuni, Hidayat, Saputra, Murti, Rachman  \& Yatini}{Priyatikanto et~al.}{2023}]{priyatikanto2023characterization}
Priyatikanto R.,  Mumpuni E.~S.,  Hidayat T.,  Saputra M.~B.,  Murti M.~D.,  Rachman A.,   Yatini C.~Y.,  2023, Monthly Notices of the Royal Astronomical Society, 518, 4073

\bibitem[\protect\citeauthoryear{Pullen, Kapp, McCallister, Chang, Gehrels, Garzione, Heermance  \& Ding}{Pullen et~al.}{2011}]{pullen2011qaidam}
Pullen A.,  Kapp P.,  McCallister A.~T.,  Chang H.,  Gehrels G.~E.,  Garzione C.~N.,  Heermance R.~V.,   Ding L.,  2011, Geology, 39, 1031

\bibitem[\protect\citeauthoryear{Qian, Yao, Zou, Wang  \& Yin}{Qian et~al.}{2019}]{qian2019empirical}
Qian X.,  Yao Y.,  Zou L.,  Wang H.,   Yin J.,  2019, Publications of the Astronomical Society of the Pacific, 131, 125001

\bibitem[\protect\citeauthoryear{S{\'a}nchez~de Miguel, Aub{\'e}, Zamorano, Kocifaj, Roby  \& Tapia}{S{\'a}nchez~de Miguel et~al.}{2017}]{sanchez2017sky}
S{\'a}nchez~de Miguel A.,  Aub{\'e} M.,  Zamorano J.,  Kocifaj M.,  Roby J.,   Tapia C.,  2017, Monthly Notices of the Royal Astronomical Society, 467, 2966

\bibitem[\protect\citeauthoryear{Sch{\"o}ck et~al.,}{Sch{\"o}ck et~al.}{2009}]{schock2009thirty}
Sch{\"o}ck M.,  et~al., 2009, Publications of the Astronomical Society of the Pacific, 121, 384

\bibitem[\protect\citeauthoryear{{\'S}ci{\k{e}}{\.z}or \& Czaplicka}{{\'S}ci{\k{e}}{\.z}or \& Czaplicka}{2020}]{scikezor2020impact}
{\'S}ci{\k{e}}{\.z}or T.,  Czaplicka A.,  2020, Journal of Quantitative Spectroscopy and Radiative Transfer, 254, 107168

\bibitem[\protect\citeauthoryear{Seidel, Otarola  \& Th{\'e}ron}{Seidel et~al.}{2023}]{seidel2023impact}
Seidel J.~V.,  Otarola A.,   Th{\'e}ron V.,  2023, Atmosphere, 14, 1511

\bibitem[\protect\citeauthoryear{Vernin et~al.,}{Vernin et~al.}{2011}]{vernin2011european}
Vernin J.,  et~al., 2011, Publications of the Astronomical Society of the Pacific, 123, 1334

\bibitem[\protect\citeauthoryear{Wallner \& Kocifaj}{Wallner \& Kocifaj}{2023}]{wallner2023aerosol}
Wallner S.,  Kocifaj M.,  2023, Journal of Environmental Management, 335, 117534

\bibitem[\protect\citeauthoryear{Wang}{Wang}{1963}]{wang1963preliminary}
Wang S.,  1963, Acta Meteor. Sinica, 33, 361

\bibitem[\protect\citeauthoryear{Wang}{Wang}{2019}]{wang2019three}
Wang C.,  2019, Climate Dynamics, 53, 5119

\bibitem[\protect\citeauthoryear{Wang, Wu  \& Fu}{Wang et~al.}{2000}]{wang2000pacific}
Wang B.,  Wu R.,   Fu X.,  2000, Journal of climate, 13, 1517

\bibitem[\protect\citeauthoryear{Wang, Zhou, Xing, Tang, Ma  \& Ding}{Wang et~al.}{2019}]{wang2019comparison}
Wang Z.,  Zhou X.,  Xing Z.,  Tang Q.,  Ma D.,   Ding C.,  2019, Theoretical and Applied Climatology, 137, 1541

\bibitem[\protect\citeauthoryear{Wang, Zhao, Sun, Yang, Deng, He, Rong  \& Wei}{Wang et~al.}{2024}]{wang2024phone}
Wang Y.,  Zhao Y.,  Sun W.,  Yang F.,  Deng L.,  He F.,  Rong Z.,   Wei Y.,  2024, Publications of the Astronomical Society of the Pacific, 136, 044501

\bibitem[\protect\citeauthoryear{Welch \& Tekatch}{Welch \& Tekatch}{2016}]{sqm2016}
Welch D.,  Tekatch A.,  2016, Sky quality meter – LE

\bibitem[\protect\citeauthoryear{Wen, Ke-li  \& Hao}{Wen et~al.}{2004}]{wen2004influences}
Wen F.,  Ke-li W.,   Hao J.,  2004, Plateau Meteorology, 23, 271

\bibitem[\protect\citeauthoryear{Wu, Zhou  \& Li}{Wu et~al.}{2017a}]{wu2017atmospheric(a)}
Wu B.,  Zhou T.,   Li T.,  2017a, Journal of Climate, 30, 9621

\bibitem[\protect\citeauthoryear{Wu, Zhou  \& Li}{Wu et~al.}{2017b}]{wu2017atmospheric(b)}
Wu B.,  Zhou T.,   Li T.,  2017b, Journal of Climate, 30, 9637

\bibitem[\protect\citeauthoryear{Wu, Ding, Liu  \& Li}{Wu et~al.}{2019}]{wu2019characteristics}
Wu P.,  Ding Y.,  Liu Y.,   Li X.,  2019, International Journal of Climatology, 39, 5241

\bibitem[\protect\citeauthoryear{Yeh et~al.,}{Yeh et~al.}{2018}]{yeh2018enso}
Yeh S.-W.,  et~al., 2018, Reviews of Geophysics, 56, 185

\bibitem[\protect\citeauthoryear{Yihui et~al.,}{Yihui et~al.}{2023}]{ding2023regional}
Yihui D.,  et~al., 2023, Advances in Earth Science, 38, 551

\bibitem[\protect\citeauthoryear{Yu, Wang  \& Li}{Yu et~al.}{2003}]{yu2003spatial}
Yu Y.,  Wang J.,   Li Q.,  2003, Journal of Glaciology and Geocryology, 25, 149

\bibitem[\protect\citeauthoryear{Zhang, Wallace  \& Battisti}{Zhang et~al.}{1997}]{zhang1997enso}
Zhang Y.,  Wallace J.~M.,   Battisti D.~S.,  1997, Journal of climate, 10, 1004

\bibitem[\protect\citeauthoryear{Zhao et~al.,}{Zhao et~al.}{2022}]{zhao2022long}
Zhao Y.,  et~al., 2022, Astronomy \& Astrophysics, 663, A34

\bibitem[\protect\citeauthoryear{Zhu, Zhang, Sun, Li, Yang, He, Weng  \& Deng}{Zhu et~al.}{2023}]{zhu2023astronomical}
Zhu L.,  Zhang H.,  Sun G.,  Li X.,  Yang F.,  He F.,  Weng N.,   Deng L.,  2023, Monthly Notices of the Royal Astronomical Society, 522, 1419

\makeatother
\end{thebibliography}






\bsp	
\label{lastpage}
\end{document}